\documentclass[journal = mamobox, manuscript=article]{achemso}
\usepackage[utf8]{inputenc}
\usepackage[english]{babel}
\usepackage{graphicx}
\usepackage{amsmath}
\usepackage{amssymb}
\usepackage{textcomp}
\usepackage{hyperref}
\usepackage{tabularx}
\usepackage{multirow}
\usepackage{sidecap}
\usepackage{wrapfig}
\usepackage{subcaption}
\usepackage{hyperref}
\usepackage{comment}
\usepackage{microtype}
\usepackage[numbers]{natbib}
\usepackage{natmove}
\usepackage{hyperref}
\usepackage{listings}

\lstdefinestyle{mystyle}{
    backgroundcolor=\color{backcolour},
    commentstyle=\color{codegreen},
    keywordstyle=\color{magenta},
    numberstyle=\tiny\color{codegray},
    stringstyle=\color{codepurple},
    basicstyle=\ttfamily\footnotesize,
    breakatwhitespace=false,
    breaklines=true,
    captionpos=b,
    keepspaces=true,
    numbers=left,
    numbersep=5pt,
    showspaces=false,
    showstringspaces=false,
    showtabs=false,
    tabsize=2
}

\usepackage{color}
\usepackage{colortbl}
\usepackage{xcolor}
\usepackage{csquotes}
\usepackage[nolist,nohyperlinks]{acronym}

\begin{acronym}[RNEMDS]
  \acro{ANN}{artificial neural network}
  \acro{XAI}{explainable AI}
  \acro{BCP}{block copolymer}
  \acro{COM}{center of mass}
  \acro{CPU}{central processing unit}
  \acro{DSA}{directed self-assembly}
  \acro{DDP}{Distributed Data Parallelization}
  \acro{EUV}{extreme ultraviolet lithography}
  \acro{FFT}{Fast Fourier Transform}
  \acro{GAN}{generative adversarial networks}
  \acro{WGAN}{Wasserstein-GAN}
  \acro{GPGPU}{general purpose computing on graphics processing units}
  \acro{GPU}{graphics processing unit}
  \acro{HDF5}{hierarchical data format version 5}
  \acro{HPC}{high performance computing}
  \acro{LSTM}{long-short term memory neural network}
  \acro{MC}{Monte-Carlo}
  \acro{MCS}{Monte-Carlo steps}
  \acro{MD}{molecular dynamics}
  \acro{MFA}{mean-field approximation}
  \acro{ML}{machine learning}
  \acro{MSD}{mean squared displacement}
  \acro{ODT}{order-disorder-transition}
  \acro{PMMA}{poly(methyl methacrylate)}
  \acro{PS}{poly(1-phenylethene)}
  \acro{SCFT}{self-consistent field-theory}
  \acro{SCMF}{single-chain-in-mean-field}
  \acro{SMC}{smart Monte-Carlo}
  \acro{SOMA}{SOft coarse grained MC Acceleration}
  \acro{UNet}{UNet neural network architecture}
\end{acronym}

\newcommand{\Tr}{T_{\textrm R}}

\newlength{\figurewidth}
\newlength{\multifigurewidth}
\title{Prediction of Diblock Copolymer Morphology via Machine Learning}
\author{Hyun Park}
\affiliation[UIUC]{Theoretical and Computational Biophysics Group, NIH Resource Center for Macromolecular Modeling and Bioinformatics, Beckman Institute for Advanced Science and Technology, University of Illinois at Urbana-Champaign, Urbana, Illinois 61801, USA}
\alsoaffiliation[BIOP]{Center for Biophysics and Quantitative Biology, University of Illinois at Urbana-Champaign, Urbana, Illinois 61801, USA}
\alsoaffiliation[ARG]{Data Science and Learning Division, Argonne National Laboratory, Lemont, Illinois 60439, USA}
\altaffiliation{Contributed equally to this work}

\author{Boyuan Yu}
\affiliation[PME]{Pritzker School of Molecular Engineering, University of Chicago, 5640 Ellis Ave, Chicago, Illinois 60637, USA}
\altaffiliation{Contributed equally to this work}

\author{Juhae Park}
\affiliation[PME]{Pritzker School of Molecular Engineering, University of Chicago, 5640 Ellis Ave, Chicago, Illinois 60637, USA}

\author{Ge Sun}
\affiliation[PME]{Pritzker School of Molecular Engineering, University of Chicago, 5640 Ellis Ave, Chicago, Illinois 60637, USA}

\author{Emad Tajkhorshid}
\affiliation[UIUC]{Theoretical and Computational Biophysics Group, Beckman Institute for Advanced Science and Technology, University of Illinois at Urbana-Champaign, Urbana, Illinois 61801, USA}
\alsoaffiliation[BIOP]{Center for Biophysics and Quantitative Biology, University of Illinois at Urbana-Champaign, Urbana, Illinois 61801, USA}
\alsoaffiliation[BIOC]{Department of Biochemistry, University of Illinois at Urbana-Champaign, Urbana, Illinois 61801, USA}

\author{Juan J. de Pablo}
\affiliation[PME]{Pritzker School of Molecular Engineering, University of Chicago, 5640 Ellis Ave, Chicago, Illinois 60637, USA}
\email{depablo@uchicago.edu}

\author{Ludwig Schneider}
\affiliation[PME]{Pritzker School of Molecular Engineering, University of Chicago, 5640 Ellis Ave, Chicago, Illinois 60637, USA}
\email{ludwigschneider@uchicago.edu}


\begin{document}

\maketitle

\begin{abstract}
A machine learning approach is presented to accelerate the computation of block polymer morphology evolution for large domains over long timescales. The strategy exploits the separation of characteristic times between coarse-grained particle evolution on the monomer scale and slow morphological evolution over mesoscopic scales. In contrast to empirical continuum models, the proposed approach learns stochastically driven defect annihilation processes directly from particle-based simulations. A UNet architecture that respects different boundary conditions is adopted, thereby allowing periodic and fixed substrate boundary conditions of arbitrary shape. Physical concepts are also introduced via the loss function and symmetries are incorporated via data augmentation. The model is validated using three different use cases. Explainable artificial intelligence methods are applied to visualize the morphology evolution over time. This approach enables the generation of large system sizes and long trajectories to investigate defect densities and their evolution under different types of confinement. As an application, we demonstrate the importance of accessing late-stage morphologies for understanding particle diffusion inside a single block. This work has implications for directed self-assembly and materials design in micro-electronics, battery materials, and membranes.
\end{abstract}

\section{Introduction}

Diblock copolymers have the ability to form nano-structured materials through microphase separation.
Their equilibrium bulk properties\cite{matsen2001standard} are well understood, but questions remain about their dynamics and morphological evolution over long time scales.
Further research into their dynamics is necessary to develop emerging applications like battery materials\cite{mayer2021block,sharon2021molecular,schneider2019engineering,shen2018diffusion} and micro-electronic device fabrication. This issue is particularly critical in the context of the kinetics of  \ac{DSA}\cite{stoykovich2007directed,kagan2020self,feng2022optimized,cheng2004nanostructure,ruiz2008density,darling2007directing,bita2008graphoepitaxy,tang2008evolution,luo2013directed,bates2014block,hu2014directed,morris15directed}.
\ac{DSA}, where copolymers are of interest for pattern rectification in nano-lithography \cite{chang2022sequential} \cite{liu2013chemical}.
Note that with the advent of \ac{EUV} for micro-electronics fabrication, the removal of defects in target lamellar structures has become increasingly important\cite{gronheid2013rectification,guo2019lcdu,chi2017electrical}.

For battery materials, micro-electronic and other applications, it is important to understand not only the evolution of the bulk morphology of diblock copolymers, but also their interaction with boundaries.
Boundaries can affect the final equilibrium morphology\cite{huang1998using,tsori2000defects} of these materials.
For example, symmetric diblock copolymers that form lamellae will generate ''standing'' lamellae on a neutral surface, but will form ''parallel'' lamellae on an attractive surface.
More complex geometries can be conceived and exploited to design specific structures\cite{michman2019controlled,schneider2020symmetric}.
Research in this area has the potential to enable controlled, high-precision manufacturing of materials for a range of applications.

Designing confinements to achieve a specific polymer morphology can be an effective means to engineer fabrication processes.
For example, contact holes can be filled with a polymer morphology to create smaller holes with a defined size\cite{peters2015graphoepitaxial,stoykovich2007directed,zhou2018studying}.
In order to iteratively adjust boundary conditions to achieve a desired final conformation, it is important to be able to rapidly predict the final morphology from knowledge of the boundary conditions.
Computer simulations are a valuable tool for this purpose, but particle-based simulations can be challenging and resource-intensive\cite{muller2011studying,muller2011speeding}.
Coarse-grained models such as the \ac{SCMF} model can enable large-scale simulations of morphological  evolution\cite{daoulas2006single,schneider2018multi,schneider2019engineering}, but are not suitable for regular, large-scale production.

In recent decades, continuum models have been developed to predict the morphology and kinetics of diblock copolymers\cite{muller2018continuum,rottler2020kinetic,grest1996efficient,dreyer2022simulation,swift1977hydrodynamic,fredrickson1987fluctuation,fredrickson1989kinetics,elder1992ordering,seul1995domain,tsori2001diblock,ohta1986equilibrium,kawasaki1988equilibrium,ohta1993anomalous,ren2001cell,choksi2003derivation,lequieu2023combining}.
These models do not track the motion of individual particles, but instead use a spatially resolved order parameter $\varphi(\boldsymbol{r})$ to describe the composition of the system.
This allows for a solution of the equation of motion for the order parameter only, which is usually discretized on a grid, thereby providing computational advantages.
However, these models lack the particle-level detail that is often of interest for materials design, a shortcoming that can make it challenging to predict kinetics accurately.
Field theoretic models have been able to address this challenge \cite{rottler2020kinetic,lennon2008numerical,delaney2016recent}, but involve additional mathematical and computational complications.

In recent years, the use of \ac{ML} in combination with scientific models has been explored to address challenging or resource intense questions in science\cite{jumper2021highly,unke2021machine}.
In the context of diblock copolymers, \ac{ML} has been used to improve our understanding of the phase behavior in these systems\cite{arora2021random,zhao2021autonomous,aoyagi2021deep}.
In particular, we have introduced an approach~\cite{schneider2021combining} that uses a dataset from particle-based simulations to train a neural network to predict the late-stage morphological evolution of a diblock copolymer using large time steps.
However, that approach~\cite{schneider2021combining} was limited to two dimensions and implicit boundary conditions, and was primarily focused on understanding defect annihilation in thin film.

In this work, we significantly improve the underlying \ac{ML} approach to allow for investigations in three dimensions and interactions with explicit boundary conditions.
The boundary conditions are not limited to the edges of a cubic box, but can take any shape that can be resolved by the underlying grid.
This enables us to study the influence of complex boundary shapes on the late-stage morphology evolution of copolymers in three dimensions.
The \ac{ML} scheme accelerates the integration, eliminating time constraints and making it possible to iteratively optimize boundary conditions until a desired structure or property is achieved.

The manuscript is organized as follows:
in \nameref{sec:methods}, we provide details about the particle-based simulations and the creation of the training dataset.
We also describe the \ac{ML} architecture that is deployed to learn the kinetics of the morphology evolution.
In \nameref{sec:results}, we discuss how the \ac{ML} model was trained and verify that the predictions are physically reasonable.
We then investigate model systems with interesting boundary morphologies and study the resulting kinetics over long time scales.

\section{Simulation methods}\label{sec:methods}
We introduce a methodology aimed at decoupling the slow time scales of morphology evolution from the fast time scales of molecular motion.
This decoupling is possible because these time scales are sufficiently different. Our methodology focuses on the post-spinodal decomposition stage, where the interaction annihilation of defects in the lamellar system dominates the dynamics.
Specifically, we aim to overcome the elevated free-energy barriers between grains of distinct orientations with divergent lamellae orientations in three-dimensional space.

To this end, we propose to build a Markov chain that predicts the temporal evolution of the morphology from two configurations at different time points, namely, $\varrho_{t-\Delta t}$ and $\varrho_t$ at time $t$, to a later time $t+\Delta t$. Although this approach is not strictly Markovian, since it uses two past time steps for prediction, we reinterpret our state as the combination of $(\varrho_{t-\Delta t}, \varrho_t)$ that is advanced to $(\varrho_t, \varrho_{t+\Delta t})$, making the prediction Markovian again. Notably, we learn the dynamics of the Markov chain from a dataset of particle-based simulations, rather than relying on empirical continuum models. This approach effectively decouples polymer motion from morphological change.

\begin{align}
    \varrho_{t+\Delta t} = \mathcal{T}(\varrho_{t-\Delta t} ,\varrho_t| \theta)
\end{align}
The proposed methodology involves using machine learning to learn the weights $\theta$ for the time evolution operator $\mathcal{T}$, assuming that it is time-independent long after spinodal decomposition has occurred. We acknowledge that morphology evolution is much faster during spinodal decomposition and is highly dependent on the underlying particle dynamics\cite{rottler2020kinetic}.
We also assume that the system can be described by its spatial composition $\varphi(\boldsymbol{r})$, which neglects all particle positions. A simpler version of this proposed framework has been shown to work in two dimensions\cite{schneider2021combining}, and here we expand it to bulk, three dimensional systems, as well as the effects of complex confining geometries, such as substrates and chemical patterns for \ac{DSA}.

In the subsequent section \nameref{sec:data-gen}, we describe how we generate a dataset to learn the time evolution operator $\mathcal{T}$ using a coarse-grained particle-based model through \ac{ML}.

An alternative approach that has been considered in the literature is to generate equilibrium morphologies directly from random inputs using \ac{GAN} or \ac{WGAN} models\cite{weng2019gan}. We also attempted this approach, but found it challenging to build satisfactory models that can incorporate non-trivial boundary conditions.
Additionally, it was also challenging to meaningfully control the input of such an approach to achieve specific points in chemical space. We therefore opted for the time evolution approach, which provides information not only about the final equilibrium configuration but also about the time evolution and how this morphology can be kinetically accessed. It is noteworthy that the kinetic approach constructs a Markov chain model, which incrementally improves the configuration until a fixed point, i.e. until a local free-energy minimum is reached, similar to diffusion-type machine learning models\cite{ho2020denoising}.
The enhanced kinetics process is applied iteratively to rectify defects, in contrast to noise reduction in machine learning diffusion models.

\subsection{Data generation with a particle-based model}\label{sec:data-gen}
In this study, we employ a soft, coarse-grained model to simulate the self-assembly and defect kinetics of diblock copolymers. Specifically, we use the \ac{SOMA} software \cite{schneider2018multi}, in which Monte Carlo simulations of a coarse-grained particle-based model are combined with the single-chain-in-mean-field (SCMF) approach \cite{daoulas2006single}. This \ac{GPU}-accelerated software enables us to investigate polymeric systems at large scales.

In our model, polymer chains are represented as discretized Gaussian chains composed of $N$ coarse-grained beads. Without loss of generality, we adopt a discretization of $N = 32$ for diblock copolymer molecules and choose symmetric lamellae-forming diblocks with $N_A = N_B = 16$. The bonded energy at a given temperature $T$ is determined by the following equation:
\begin{equation}
\frac{H_b}{k_BT}=\sum_{m}\sum_{i}\frac{3(N-1)}{2{R_{e,0}}^2}(\boldsymbol{r}_{m,i}- \boldsymbol{r}_{m,i+1})^2
\end{equation}
Here, $\boldsymbol{r}_{m,i}$ and $\boldsymbol{r}_{m,i+1}$ denote the positions of a particle and its bonded neighbor in chain m, $N$ represents the number of beads in the chain, $k_B$ represents Boltzmann constant, and ${R_{e,0}}^2$ denotes the mean-squared end-to-end distance of an ideal chain.

The non-bonded interactions between the coarse-grained beads are described by an excess free energy functional, $H_{nb}[\phi_A, \phi_B]$, evaluated on a grid as a function of the local densities $\phi_{\alpha}(r)$, with $\alpha \in \{A, B\}$. The soft, non-bonded interactions are given by:

\begin{equation}
\frac{H_{nb}[\hat{\phi}_{\alpha}]}{k_BT} = \frac{\rho_0\Delta{L}^3}{N}\sum_{c\in{cell}}\left(\frac{\kappa_0N}{2}(\hat{\phi}_A(c)+\hat{\phi}_B(c)-1
)^2-\frac{\chi_0N}{4}(\hat{\phi}_A(c)-\hat{\phi}_B(c))^2\right)
\end{equation}

Here, $\rho_0$ represents the average particle density of a homogeneous system, and $\Delta{L}$ is the grid cell size, set to $\Delta{L} = R_e/6$ in this work. Here, $\kappa_0$ is related to the inverse isothermal compressibility, which constrains particle-density fluctuations, and $\chi_0$ is the Flory-Huggins interaction parameter, which measures the thermodynamic incompatibility of different species. For this study, we select as model parameters $\kappa_0N = 30$, which effectively restrains density fluctuations, and $\chi_0N = 20$, which leads to weak segregation in lamellar microstructures.
Here, $\hat{\phi_{\alpha}}$ represents a normalized particle density given in grid cell $c$ by particle coordinates {$\mathbf{r}_j$}, and is written as:

\begin{equation}
\hat{\phi}_{\alpha}(c) = \frac{1}{\Delta L^3\rho_0}\sum_{j\in{beads}}\Pi_c(\mathbf{r}_j)\delta_{type(j), i}
\end{equation}

where the summation runs over all beads, and the assignment function, $\Pi_c(\mathbf{r}_j)$, is a characteristic function of the grid cell, $\rho_0$ is the average particle number density, and $\Delta L$ denotes the linear spacing of a grid cell.

The simulations were initialized by randomly placing $N$ beads with a Gaussian distribution within a cubic box of dimensions $L_x \times L_y \times L_z = 10 R_{e,0} \times 10 R_{e,0} \times 10 R_{e,0}$. The repulsive interaction between the two blocks was instantaneously quenched to $\chi_{0}N = 20$ at the start of the simulation. Local normalized densities ($\varphi$) were recorded every 50000 \ac{MC} cycles (equivalent to $5 \Tr$), where $\Tr=R_e^2/D$ is the molecule relaxation time in which a polymer diffuses its own extension in a homopolymer bulk and is related to the Rouse time. The local normalized density in each grid cell was calculated using the formula $\varphi = \phi_A/(\phi_A + \phi_B)$. The values of $\varphi$ ranged from 0 to 1, where $\varphi = 0$ corresponds to a grid cell containing only component B, $\varphi = 0.5$ represents an interface between the A and B components, and $\varphi = 1$ corresponds to a grid cell containing only component A.

\subsubsection{Preparation of sytems of interest}

We generated training and testing data for three distinct systems: a 3-D bulk system with periodic boundary conditions as a reference, a flat-wall system to study the effects of boundary conditions in a controlled environment, and a spherical, convex wall system, which is the most realistic configuration to include curvature and demonstrate the generalization to complex situations. In the bulk system, periodic boundary conditions were used in the x, y, and z directions. For the flat-wall system, periodic boundary conditions were employed in the y and z directions, and impenetrable walls were implemented in the x direction to simulate a film structure. In the system with spherical walls, an impenetrable solid sphere was placed in the center of the box with a given radius ($r = 5R_e$), and periodic boundary conditions were adopted in the x, y, and z directions. To train the system with impenetrable walls, we assigned $\varphi = -1$ for the substrate.

We generated 1024, 1000, and 612 samples for the 3-D bulk, flat-wall, and spherical-wall systems, respectively. The initial configurations were randomly generated, and a different initial seed for the \ac{MC} algorithm was used for each simulation. Each sample could be simulated within one hour on a single Nvidia A100 GPU.

\subsubsection{Data augmentation}\label{sec:aug}
Data augmentation is a crucial technique for enhancing the accuracy and robustness of neural network models \cite{yang2022image}. It involves applying transformations to the original dataset, such as flipping, color jittering, and cropping, to not only increase the number of available items but also improve the model's performance \cite{geirhos2018imagenet}.

However, in our case, we are dealing with physical data that represents polymer composition density and, as such, data augmentation must be restricted to maintain the physical interpretation of the results. We observe that the data have smooth neighboring voxel densities and local mass conservation principles, and also have periodic boundary conditions. Therefore, we limit our data augmentation to flips and rotations only, which we have found to improve prediction quality.

Our flips (one of the two data augmentation mentioned above) involve reversing the order of elements along an axis, while rotations (the other data augmentation) entail rotating by multiples of $90^{\circ}$ with respect to a plane axis. We randomly select axis and plane axis for both augmentation methods during training, and refrain from applying them during inference.

Since morphology evolution is rotationally and translationally symmetric, we ensure that our model learns an evolution process that respects these symmetries through data augmentation. The discrete simulation box we work with implies that certain angles of lamellar orientation are favored due to commensurability with the periodic simulation box. Nonetheless, we observe that with our data augmentation approach, the final lamellar orientations are determined by the initial conditions and are not biased towards any particular orientation.

\subsection{Machine learning and simulation methods}\label{sec:ml}
In this section, we provide a detailed description of the \ac{ANN} model used to predict future spatiotemporal ($\varphi(r)$) evolution.

Our \ac{ML} architecture is based on the 3D version of the well-known \texttt{\ac{UNet}} \cite{oktay2018attention}, which was originally developed for image and object segmentation. The \texttt{\ac{UNet}} model consists of encoder and decoder components. The encoder compresses the input image size, while the decoder expands the compressed image back to the original size. Importantly, the decoder also takes into account the input image and its hidden representations from the encoder by concatenating or adding them. This ensures a good flow of information from input to output and makes the \texttt{\ac{UNet}} a good choice for image segmentation tasks. It is worth noting that there are several varieties of \texttt{\ac{UNet}} architectures optimized for different tasks, such as medical image segmentation \cite{kolavrik2019optimized} and cell detection in microscopy images \cite{futrega2022optimized}.

\begin{figure}[ht]
\centering
\includegraphics[width=\figurewidth]{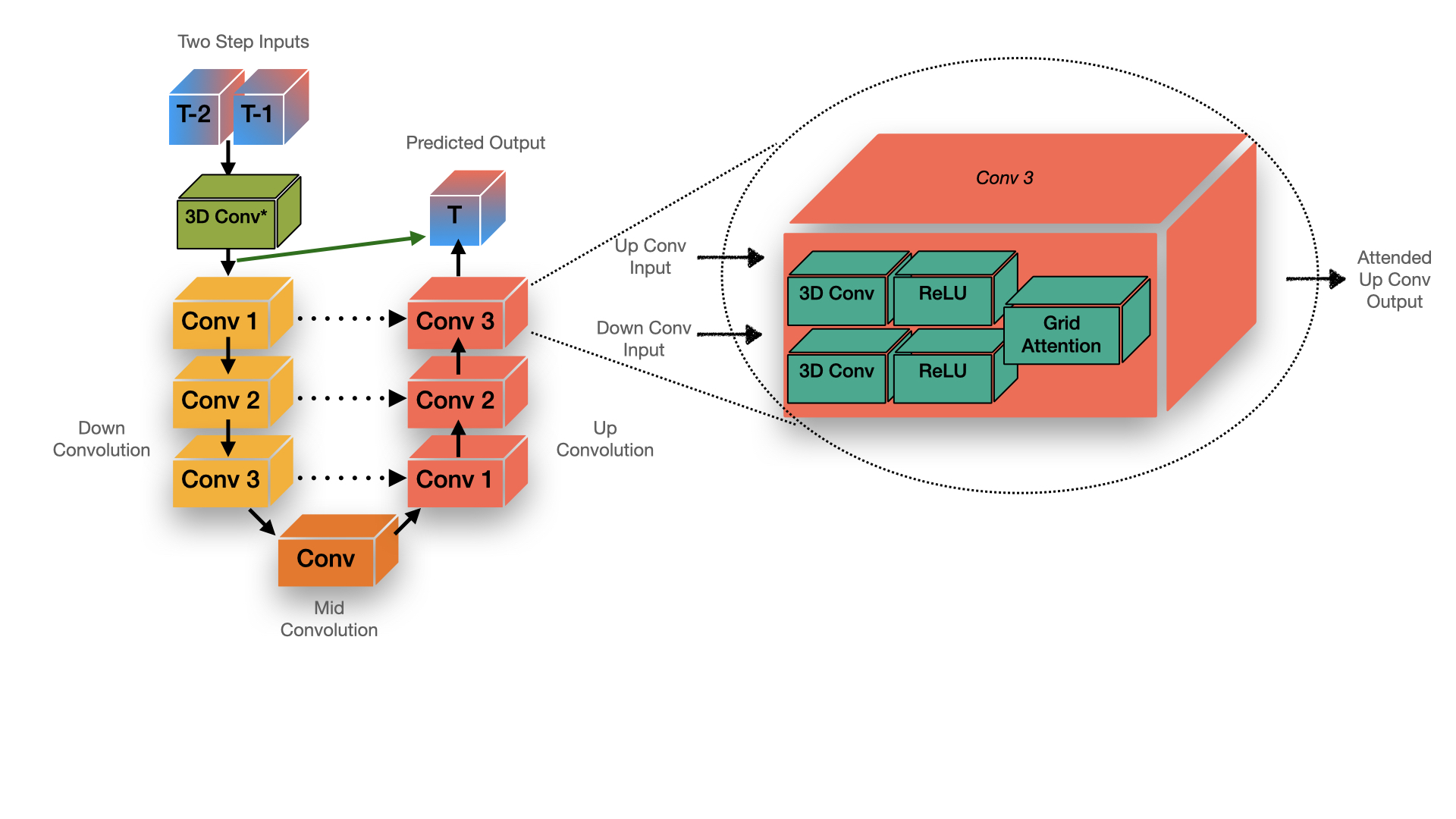}
\caption{\texttt{3D-UNet} Architecture.  Once data from two previous steps $\varphi(r)$ are passed to the encoder part of \texttt{UNet}, the input's hidden representations are extracted (i.e., dotted arrows from yellow convolution blocks to red convolution blocks) and trickle down to the next convolution unit (i.e., solid arrows between \textit{conv1 and conv2} and \textit{conv2 and conv3}). Down convolution is an encoder whose hidden representation dimension of an input becomes smaller while up convolution is a decoder whose hidden representation is add/concatenated with the encoder counterpart. This addition/concatenation leads to good mixing of information from the input to the output, i.e., the next time step predicted $\varphi(r)$. Each convolution block contains multiple convolution units, activation functions and grid attention components. The customized convolution unit, labeled \texttt{Conv*} is located at the beginning of the down convolution and the meta data (e.g., indices of masking value locations) are passed to predicted output denoted by the green arrow. Meta data are used to ensure masking values are not propagated during training and inference. The \texttt{Conv} indicates a normal convolution unit. See \autoref{fig:mask_aware_pad} for additional details about the custom convolution.}
\label{fig:attention_umap}
\end{figure}

Our work repurposes the \texttt{\ac{UNet}} architecture, as shown in Figure \ref{fig:attention_umap}, to predict the future one-step ahead $\varphi(r)$ evolution by taking in the previous two steps of $\varphi(r)$ data as input, which we designate as a Markovian process (see \nameref{sec:methods}). Since the $\varphi(r)$ data spans across 3 spatial dimensions and 1 channel dimension, our dataset is of shape \texttt{channel x dimension 1 x dimension 2 x dimension 3}, where the channel has a value 1, at a given time $t$. This enforced channel dimension exists because a convolution operator used in \texttt{UNet} requires both space and channel dimensions.

As mentioned above, our dataset consists of $\varphi(r)$ information with four dimensions: three spatial dimensions representing the indices of the voxels in grid space (a.k.a. spatial dimension data), and one channel dimension representing the density value per voxel. To predict the future one-step-ahead $\varphi(r)$ evolution, we repurpose the \texttt{\ac{UNet}} architecture as a time series prediction model. The U-shape of the \texttt{\ac{UNet}} ensures a good flow of information from the encoder to the decoder, and the input and output spatial dimension sizes are the same. The input and output shape matching for the spatial dimensions is why \texttt{\ac{UNet}} is a good choice over many other \ac{ML} architectures, without a need for further alterations. By utilizing the encoder input and the decoder output efficiently, the next time step in the morphological evolution can be predicted.

To further improve the quality of our predictions, we employ an attention mechanism proposed in Ref.~\cite{oktay2018attention}, which has been shown to improve physically relevant metrics such as volume fraction ($\Bar{\varphi}$). Our attention mechanism is a grid-attention, where each voxel has an importance value between 0 to 1. The encoder output serves as the key, providing input data representation through convolutional operations. After the decoder's output (a.k.a. query or gating vector in Ref.~\cite{oktay2018attention}, which provides summary and contextual understanding of the input data) is interpolated to match the shape of the encoder's output, both outputs are added and passed through a sigmoid activation to produce the attention map. The grid-attention differs from the traditional attention mechanism \cite{vaswani2017attention} in two ways. First, it uses a sigmoid activation instead of softmax, making each voxel's attention value spatial information dependent (in contrast to global information dependent), which prevents the attention map from producing sparse values. Second, the additive mechanism between the encoder and the decoder's output is chosen over the multiplicative mechanism because of its reportedly better performance in voxel segmentation tasks in ~\cite{oktay2018attention}.

The attention map calculated can be used for forward propagation, backward propagation and \ac{XAI}. During the forward propagation, the attention map prunes input features so that only relevant activation values are retained. During backward propagation, the attention map regulates how much gradient of the input representations should be updated, which allows \texttt{\ac{UNet}} parameters to be updated for only relevant spatial regions. For \ac{XAI}, we can utilize the attention map to understand what our \ac{ML} model focuses on to annihilate defects during the morphology evolution. The details of forward and backward propagation are available in the literature\cite{oktay2018attention}; the details of \ac{XAI} are provided in \nameref{sec:xai}.

In the original formulation of convolutional neural network based \ac{ML} models, batch normalization is used to train faster and with lower errors and better stability\cite{ioffe2015batch}\cite{bjorck2018understanding}.
Batch normalization calculates batch statistics of the input batch during training, which is used for standardizing input values (i.e. zero mean with standard deviation 1). This approach has proven to reduce exploding and fluctuating gradients of the underlying parameters.
During inference, the running mean and standard deviation are used, instead of per-batch statistics for a new input.
One caveat is that a large batch size is required in order for batch normalization to work well. In our work, a batch size of 32 is the highest available, due to the memory requirements of our data.
Each input is 3-dimensional, with input size 64 in each spatial dimension, plus channel dimension with size 2 (i.e. $\texttt{batch x 2 x 64 x 64 x 64}$; 2 is previous 2 time steps as an input) which is 64 times larger than the 2-dimensional (e.g. image) input with size 64 for each spatial dimension and channel dimension size 2. Therefore, due to the high memory requirements arising from large dimension of our dataset, we are limited to a small batch size (i.e. 32). However, the batch normalization with such as small value can be detrimental for stable training of the \ac{ML} model. Because the per-batch estimators (i.e. mean and standard deviation of spatial information computed from each batch to another) fluctuate considerably, the gradient of the parameters also fluctuate, leadin to unconverged parameterization of the \ac{ML} model.
Batch normalization is thus unsuitable when this normalization is used and we turn our attention to normalization techniques that are independent of batch size.
Layer\cite{ba2016layer} and group\cite{wu2018group} normalization calculate feature (a.k.a. channel) statistics, which allows for smaller batch sizes, and can then be used to standardize input.
Group normalization with a group size set to 1 (i.e. all channels are the same group) is chosen for our work, which is essentially the same as for the layer normalization.
To ensure our output is within the range of 0 to 1, because our predicted output is density value for each voxel w,e apply sigmoid activation layer on the output. We can then interpret the output as local volume fraction ($\Bar{\varphi}$) of the diblock copolymer material, which is subject to physically relevant analysis and regularization (i.e., volume fraction analysis and loss function) during training and inference.

\begin{figure}[ht]
\centering
\includegraphics[width=\figurewidth]{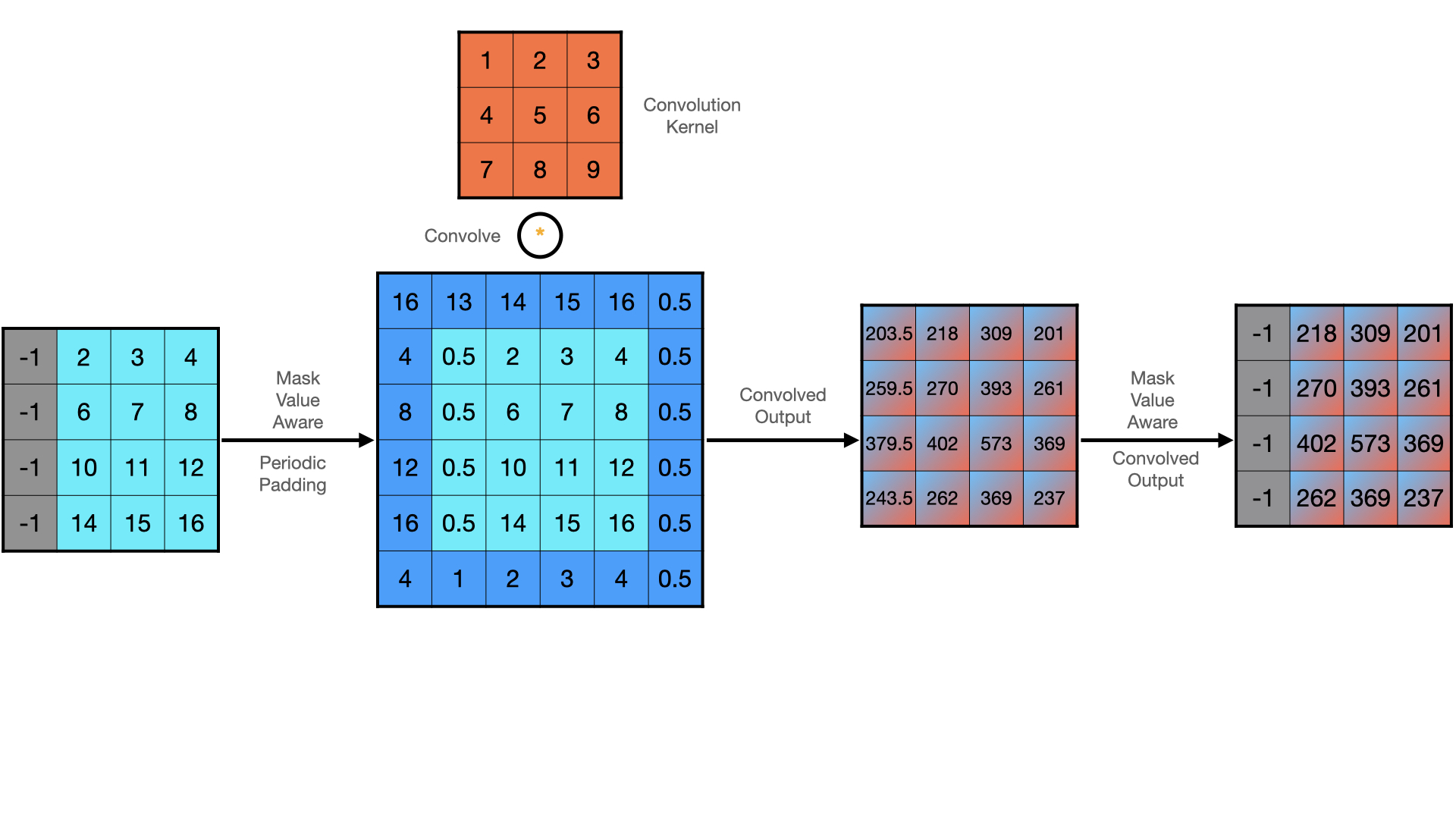}
\caption{Custom convolution with periodic padding and mask location awareness. An input (cyan with gray box) with a masking value of -1 is given to our custom convolution unit. Then the input is padded with periodicity considered (dark blue grids surrounding original cyan box). The masking value is then substituted to 0.5, an expected volume fraction (i.e., neural density value),  since propagating a negative masking value can cause instability during training. Then a learnable kernel (orange box; a.k.a. filter or convolution weight) convolves the padded input. This results in mixed information (i.e., convolution output). Finally, we place our masking value (-1) back, to ensure the mixed information does not update masking values and also to ensure correct output values (i.e., predicted next step spatial composition). }
\label{fig:mask_aware_pad}
\end{figure}

Finally, we use a custom convolution (i.e., labeled green \texttt{Conv*} in \autoref{fig:attention_umap}) in the encoder of our \texttt{\ac{UNet}} architecture to identify the substrate in the data. The locations of the substrate (i.e., masking value of -1; substrate value) are recorded as masking value indices that are substituted with a neutral density value (e.g., $1/2$). Here, we choose a neutral wall interaction $1/2$ for both blocks, but believe that the setup is generalizable to preferential surfaces $\neq 1/2$ as well.
A non-neutral setup favors the interaction with one of the two polymer blocks. These preferences can have a spatial pattern on the surface analogous to chemical patterns in applications of \ac{DSA}. The detailed mechanism of identifying masking value indices, substituting to neutral density value, and convolution, are illustrated in \autoref{fig:mask_aware_pad}.

After the output is predicted, we apply sigmoid activation and plug the masking values back into the output. This way, our output has physically correct density values except for the masking value indices. The masking value indices obtained from custom convolution at the beginning of the down convolution are used to place the masking values in the predicted spatial composition output (i.e., output of decoder), so that substrate masking value will not be updated (i.e., green arrow in \autoref{fig:attention_umap}; also see \autoref{fig:mask_aware_pad}). Since this custom convolution is integrated into our \texttt{\ac{UNet}} model, the entire process is performed end-to-end.

In addition, to improve the physically relevant metrics such as $\Bar{\varphi}$ predicted by our model, we utilize an attention mechanism proposed in Ref.~ \cite{oktay2018attention}.  Also, we use group normalization which helps training the neural network to train more stably with low error predictions of output. Lastly, we developed a custom convolution unit so that we can take advantage of masking value (i.e., location of substrates) during training and inference.

\subsubsection{Loss function}\label{sec:loss}
In our work, we introduce two physics-informed loss functions in addition to a widely used \ac{ML} loss function. The \ac{ML} loss function used is here is smooth L1 loss (sL1) and two physics-informed loss functions are the volume fraction and the structure factor peak index loss functions.

sL1 is a metric to compare two tensor element-wise values by alternating mean squared error (MSE) and mean absolute error (MAE), depending on the tensor difference and hyperparameter $\beta$. The usage of sL1 loss has been shown to prevent exploding gradient in some cases \cite{girshick2015fast}. In our work, we chose sL1 loss because ($L_{\texttt{sL1}}$)  models trained with this loss function performed  better than MSE or MAE loss trained models.

Volume fraction($\Bar{\varphi}$) is an average density of all the voxels. $\Bar{\varphi}$ value 0.5 is an average expected value of spatial composition ($\varphi(r)$) at all times during morphology evolution.
A physics-informed metric can serve as both a loss function during training and as a measure of the model's prediction quality. For instance, in our work, we employ a volume fraction loss ($L_{\Bar{\varphi}}$) term, which is added to the sL1 loss during training. However, when the dataset contains the substrate, we ignore the masking value (-1) of the substrate for calculating the $L_{\Bar{\varphi}}$.

The structure factor ($S(q)$) is a key characteristic of the micro-phase separated system and can be efficiently calculated via a \ac{FFT} of the real space density (i.e., spatial composition). A typical structure factor exhibits one main peak for lamellar structures, accompanied by several smaller peaks and troughs. We use the predicted output's Fourier transform to determine the highest peak's location during training, and this is added to the two previous losses (i.e., MSE and volume fraction losses). We call this loss function  the structure factor peak index loss ($L_{\texttt{S(q)}}$).

However, we have found that not all losses are equally important, possibly due to differences in loss value scale. Loss values with larger orders of magnitude tend to have a more significant impact on parameter gradient computation during backpropagation. Furthermore, physics-informed losses may be more important than pure \ac{ML} losses, since the \ac{ML} model must satisfy physical laws while generating desirable predictions. As a result, we assign weight coefficients of $L_{\texttt{sL1}} : L_{\Bar{\varphi}} : L_{\texttt{S(q)}} = 1 : 1.75 : 0.01$ to the loss terms, which appear to perform well in our experience.

\begin{figure}[ht]
\centering
\includegraphics[width=0.65\multifigurewidth]{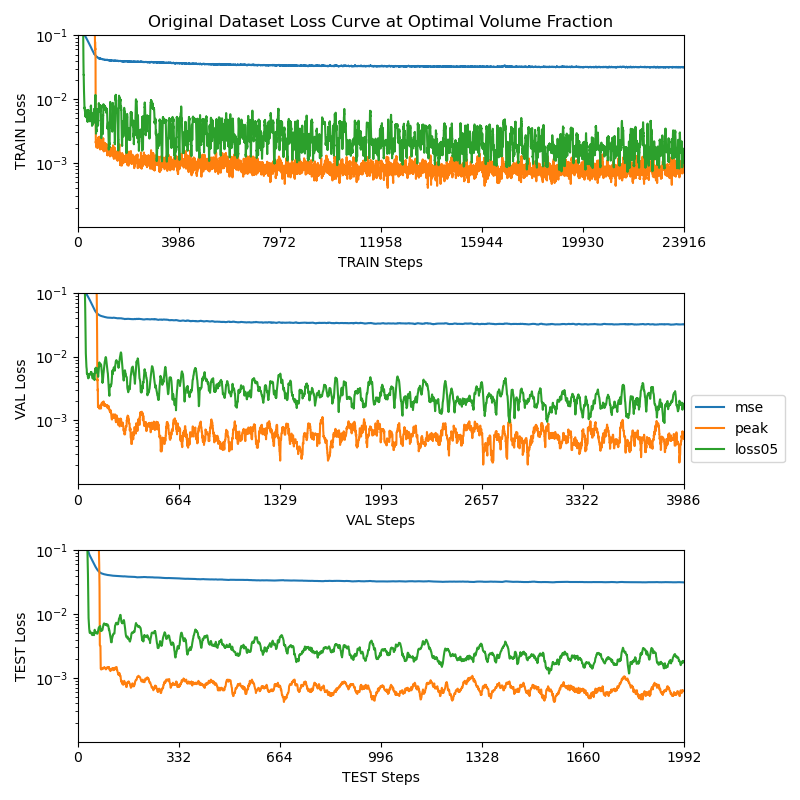}
\caption{The loss values over training/validation/test steps for bulk dataset (labeled original dataset). The top, middle and bottom panels represent train, validation and test loss curve, respectively. Three losses are used: \textit{mse} (blue), \textit{peak} (orange) and \textit{volume fraction} (green). Training steps are batch iteration steps during training to optimize our \texttt{UNet} model. The validation steps are batch iteration steps during validation to save better performing (i.e. validation loss drop) models. Lastly, the testing steps are batch iteration steps tested after all the training and saving is completed. The result is logged at every step of iteration. }
\label{fig:ori_error}
\end{figure}

The graph in \autoref{fig:ori_error} shows the time-series values of $L_{\texttt{sL1}}, L_{\Bar{\varphi}}, L_{\texttt{S(q)}}$ for the bulk dataset (labelled \textit{mse} for $L_{\texttt{sL1}}$, \textit{loss05} for $L_{\Bar{\varphi}}$ and \textit{peak} for $L_{\texttt{S(q)}}$, respectively). The top, middle, and bottom panels show train, validation, and test mode losses, respectively. We observe that all three modes converge to a small loss value early on. Similarly, \autoref{fig:flat_error} and \autoref{fig:sph_error} exhibit similar train/validation/test loss trends for simulations with flat wall and spherical wall substrates. We use exponential moving average to smooth out the loss curve for better visualization. This does not affect the analysis (e.g. convergence to a low loss value) in any way.

\section{Results and discussion}\label{sec:results}

\subsection{Training}\label{sec:train}
Training a neural network in a supervised learning fashion involves several key components, including a dataset, a model with adjustable parameters, an optimizer for updating those parameters through gradient descent, and one or more loss functions that serve as objectives to minimize during training. Hyperparameters such as learning rate and number of epochs are also important for effective training, and it is crucial to avoid overfitting the model. One common sign of overfitting is a training loss that is lower than the validation or test loss. To address this issue, we only save our models when the validation loss becomes lower than the training loss.

For our work, we trained our \ac{ML} model for up to 5000 epochs (some of the simulations were terminated sooner due to early convergence) using the AdamW optimizer \cite{loshchilov2017decoupled} and the \textit{get\_linear\_schedule\_with\_warmup} scheduler \cite{wolf-etal-2020-transformers} provided by the Hugging Face Optimizer. This scheduler gradually increases the learning rate during the warm-up stages and then linearly decreases it during training. Adjusting the learning rate has been shown to improve both training and model performance in many studies, making it a standard technique for numerous \ac{ML} tasks. As explained earlier, we set the batch size to 32, which is close to the maximum value that can be used without running into CUDA memory errors. Batch is a multiple of data points bundled together to accelerate training or inference.

During training, we utilize several loss functions, including ${L_\texttt{sL1}}$, ${L_\texttt{volume fraction}}$, and ${L_\texttt{structure factor}}$, each with different weight coefficients that were adjusted empirically to achieve optimal results. These loss functions are discussed in \nameref{sec:loss}. See \nameref{sec:Physical Evaluation} for further discussion on the effectiveness of these loss functions.
In \nameref{sec:ml}, we discussed how the grid attention mechanism \cite{oktay2018attention}, custom convolution, group normalization \cite{wu2018group}, and efficient information flow and mixing from the encoder to the decoder in the UNet architecture contribute to the high performance of our model's spatial composition prediction during morphology evolution. We have observed during both training and inference that omitting any of these components from our \ac{ML} model leads to a lesser performance.

To increase the number of samples available to our model during training, and to minimize bias in the lamellar orientation in the dataset, we augment the data via flips or rotations, as mentioned in the \nameref{sec:aug}. This leads to a model that can predict outputs with various lamellar growth patterns without a preference for a specific orientation while lowering computational demands.

In our search of the best model architecture, we evaluated several \ac{ML} models, including \texttt{\ac{UNet}}, \texttt{ConvNet} with Fourier transformation for long-range interaction \cite{schneider2021combining}, and 3D information parsing \ac{LSTM} (Eidetic 3D \ac{LSTM}; \texttt{E3D-\ac{LSTM}}) for video frame prediction \cite{wang2018eidetic}. We selected \texttt{UNet} for its consistently superior performance (i.e., good output prediction) and low computational cost (i.e., acceleration with CUDA can speed up training better). Interestingly, \texttt{\ac{UNet}} has also demonstrated its effectiveness in many state-of-the-art diffusion-based generative models \cite{ho2020denoising}, such as \texttt{DALL$\cdot$E-2} \cite{ramesh2022hierarchical} and \texttt{Imagen} \cite{saharia2022photorealistic}, which rely on it as a backbone architecture due to its good information flow from input to output.

During inference, we use the trained model to predict the output, which is then concatenated with the input to generate the next input to the model. Specifically, our model takes in two previous steps of spatial composition data (i.e., $\texttt{batch x 2 x 64 x 64 x 64}$) and predicts the next time step of morphological evolution (i.e., $\texttt{batch x 1 x 64 x 64 x 64}$). The second step of the input is then concatenated with the output in the channel dimension, which becomes the input to the model, while maintaining a consistent channel dimension of 2. Furthermore, our model can predict outputs of different spatial dimension sizes (e.g., 128 or 256 instead of 64 - see \autoref{fig:defect_bulk_large_snapshots}) due to the convolution operator being agnostic to input spatial dimension size, serving to underscore the model's generalizability with regard to input size.

We trained our model using the PyTorch \cite{NEURIPS2019_9015} framework on various machines with CUDA GPU capability. Our model can also leverage half-precision computation, which speeds up the inference steps. We also employ \ac{DDP} \cite{li2020pytorch} during both training and inference to further accelerate speed.


\subsection{Physical evaluation of the trained morphology evolution}
\label{sec:Physical Evaluation}
To ensure that our \ac{ML} models accurately predict the true defect evolution of diblock copolymers, the models must adhere to the correct underlying physical rules observed in actual \ac{MD} or \ac{MC} simulations. In this section, we assess the physical correctness of the simulation trajectories generated by our \ac{ML} models and examine how the weights of the physics-informed loss function ($L_{\Bar{\varphi}}$) affects the quality of our model's predictions. To this end, we trained models with different $L_{\Bar{\varphi}}$ weights for three systems (3D bulk, flat wall substrate, and spherical wall substrate) and generated $400$ future time (i.e., 400 $\Tr$) frames using the trained models on the test data sets.

We measured the \textit{average} volume fraction (denoted $\Bar{\varphi}_{avg}$; see \nameref{sec:3dbulk}) and the \textit{average} \ac{COM} flux (see \nameref{sec:flux} for a detailed definition) of each system averaged over the last $100$ frames and all available samples in the test data sets (\autoref{fig:evaluation}). We also calculated the \textit{average} lamellar spacing from the \ac{ML}-predicted trajectories and compared it to the lamellar spacing obtained in simulations to examine the impact of $L_{\Bar{\varphi}}$ weights on the properties of the lamellar phase (see \autoref{fig:evaluation}b, \nameref{sec:spacing} and \autoref{fig:spacing}).

\subsection{Artificial Drift Correction}
\label{sec:flux}
Although our \ac{ML} models are successful in predicting the formation of lamellar phases, some of the generated trajectories exhibit a constant motion that is not present in \ac{MC} simulations.
This is because our \ac{ML} methods do not implicitly center of mass conserve momentum.
In the particle based simulations the force-biased \ac{MC} does not promote a drift since the system is in equilibrium, but this has to be learned by the \ac{ML} model.
For instance, one of our models predicts the time evolution of a simple 3D lamellar system, and the formation of the lamellar phase is followed by the motion of the \ac{COM} along a particular direction. Therefore, when assessing model performance, the smallest drift of the system is desirable.

We define the drift as follows:
\begin{align}
\text{drift} &= \lVert\Vec{v}\rVert = \lVert(\Delta x, \Delta y, \Delta z)\rVert\\\nonumber
\Vec{v} &= (\Delta x, \Delta y, \Delta z) = \min_{\Delta x, \Delta y, \Delta z}\frac{\sum_{x,y,z = 1}^{N}[\phi(x + \Delta x, y + \Delta y, z + \Delta z, t_{curr}) - \phi(x, y, z, t_{prev})]^{2}}{N^{3}}
\end{align}
Here, $x$, $y$, and $z$ represent the index of a grid point, $t_{curr}$ represents the current time step, $t_{prev}$ represents the previous time step, and $N$ represents the number of grids along one direction.

To implicitly learn this equilibrium no drift condition, we can quantify the artificial flux in the trajectories and remove it explicitly to support the \ac{ML} with physical conservation laws.
To quantify the amount of flux, we calculate it based on two consecutive time frames.
We move the cubic density points in the current frame along a three-dimensional vector $\Vec{v} = (\Delta x, \Delta y, \Delta z)$ that minimizes the mean-square difference between the previous and moved frames.
The Euclidean norm of this vector is defined as the flux at that time step.

Since the density grid vertices are integer values (e.g., $1$ to $64$ for a simple 3D bulk system), non-integer values of $\Delta x$, $\Delta y$, or $\Delta z$ are handled as follows: we first move the system according to the integer part of the displacement, and then we calculate the new density values after moving the current frame by linear interpolation between two neighboring density points along the moving direction for decimal parts less than 1. For example, $\phi(x + \Delta x, y, z, t) = \phi(x, y, z, t) + (\phi(x + 1, y, z, t) - \phi(x, y, z, t))\Delta x$ for $0 < \Delta x < 1$.

The movement of the system also follows periodic boundary conditions. Therefore, for a flat wall substrate, we do not attempt to move the density grid along the direction that is perpendicular to the substrate. For a spherical wall substrate, we reset the density values within grids of the original sphere to $-1$ after moving the system to maintain the substrate's position.

\subsection{Lamellar spacing}
\label{sec:spacing}
To evaluate the lamellar phase properties predicted by our \ac{ML} models, we compared the lamellar spacing $L$ obtained from the \ac{ML}-predicted trajectories with the natural lamellar spacing $L_{0}$ obtained from \ac{MD} simulations. The lamellar spacing was determined using the following expression:

\begin{equation}
L = 2\pi/\lVert \vec{q}_{0} \rVert
\end{equation}

Here, $\vec{q}_{0}$ is the wave vector corresponding to the position of the first maximum in the structure factor $S(\vec{q})$, which is obtained using the \ac{FFT} of the local normalized density $\phi$. If the system includes a substrate, it is excluded from the calculation of the structure factors. The calculation of the lamellar spacing using the structure factor is well-established in the literature~\cite{vskvor2015simulation}.

\begin{figure}[ht]
\centering
\includegraphics[width=\figurewidth]{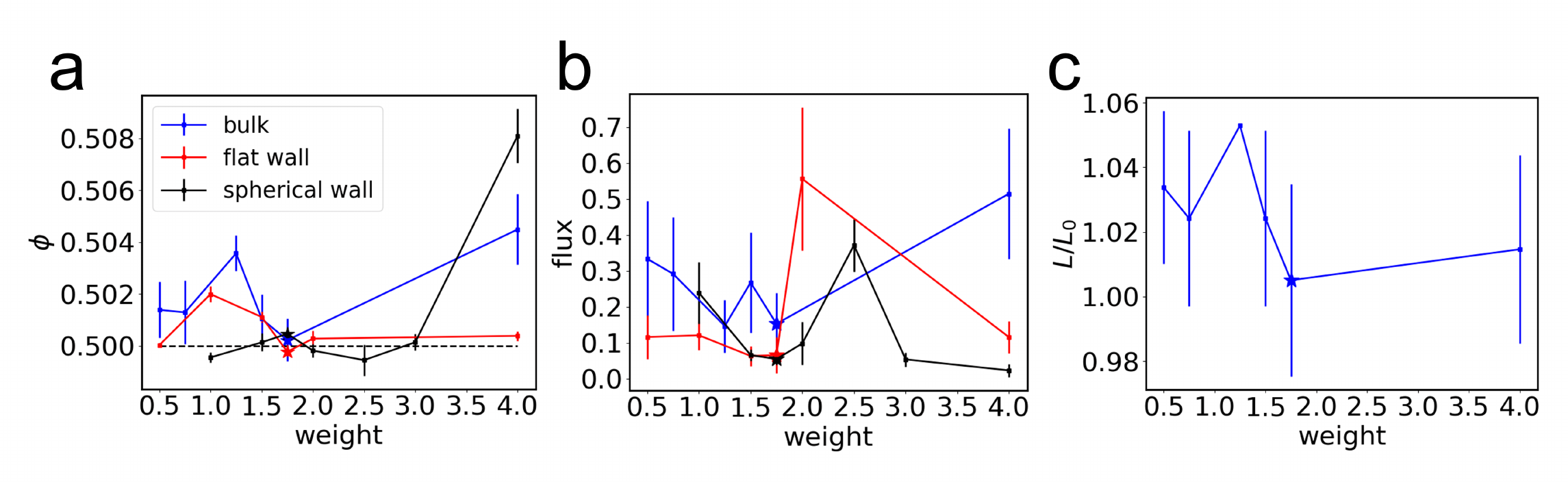}
\caption{a) \textit{Average} volume fraction $\Bar{\varphi}_{avg}$ as a function of $L_{\Bar{\varphi}}$ weight for simple bulk 3D, flat-wall substrate, and spherical-wall substrate systems. The black dashed line corresponds to the volume fraction $0.5$. b) \textit{Average} flux of system \ac{COM} as a function of $L_{\Bar{\varphi}}$ weight for simple bulk 3D, flat-wall substrate, and spherical-wall substrate systems. c) \textit{Average} lamellar spacing $L$ scaled by the natural lamellar spacing $L_{0}$ determined from simulations as a function of $L_{\Bar{\varphi}}$ weight for simple bulk 3D system. The star symbols corresponds to the weight (all of which are 1.75) of $L_{\Bar{\varphi}}$ chosen for further analysis.}
\label{fig:evaluation}
\end{figure}

\subsubsection{Reference 3D bulk system}\label{sec:3dbulk}

The \textit{average} (i.e., last 100 frame average; see \nameref{sec:Physical Evaluation}) volume fraction ($\Bar{\varphi}_{avg}$) as a function of $L_{\Bar{\varphi}}$ weight is shown by the blue curve in Figure~\ref{fig:evaluation}a, which shows a non-monotonic dependence of average volume fraction $\Bar{\varphi}_{avg}$ on loss function weight. Among the weights considered here, the value $1.75$ gives the volume fraction closest to $0.5$ indicated by the horizontal dash line. Since the diblock copolymer is symmetric in our system, the volume fraction is expected to be $0.5$. As a result, the model of $1.75$ weight of $L_{\Bar{\varphi}}$ gives the best predicted trajectories that obey the physical constraint on volume fraction.

The \textit{average} flux as a function of $L_{\Bar{\varphi}}$ weight is shown by the blue curve in Figure~\ref{fig:evaluation}b. Again, the weight values of $1.25$ and $1.75$ result in the predictions with the least amounts of flux, which is also desired for our analysis.

The \textit{average} lamellar spacing scaled by the natural lamellar spacing is plotted in Figure~\ref{fig:evaluation}c with respect to the weights of $L_{\Bar{\varphi}}$. From the figure, we can see that overall, the \ac{ML} predictions give a correct lamellar spacing, which deviates from the simulation results by not more than $6\%$. For all the weights considered here, the $1.75$ weight leads to the lamellar spacing closest to what is obtained in simulations (i.e., $L/L_{0}$ closest to $1$). This result is consistent with the analysis of average volume fraction and flux. Thus, we adopt a weight of $L_{\Bar{\varphi}}$ for simple 3D bulk systems to be $1.75$ unless mentioned otherwise.


\subsubsection{Flat-wall substrate system}

The \textit{average} volume fraction $\Bar{\varphi}_{avg}$ as a function of $L_{\Bar{\varphi}}$ weight is shown by the red curve in Figure~\ref{fig:evaluation}a. Here, the volume fraction is calculated for grid points with density value larger than $0$ (excluding the substrate). The curve shows a complex non-monotonic dependence of $\phi$ on loss function weight. Among the weights considered here, the values of $0.5$ and $1.75$ give a volume fraction close to $0.5$ (indicated by the horizontal dash line). Both the models with $0.5$ and $1.75$ weights of $L_{\Bar{\varphi}}$ give good predictions that obey the physical constraint on volume fraction.

The \textit{average} flux as a function of $L_{\Bar{\varphi}}$ weight is shown by the red curve in Figure~\ref{fig:evaluation}a. Again, the weight value of $1.75$ leads to predictions having the least amount of flux.

To estimate the \textit{average} lamellar spacing of the flat-wall substrate system from \ac{ML} predictions, we remove the substrate region from the calculation of structure factors. From the \autoref{fig:spacing}, we can see that all the weights used here yield good agreement with the simulation results (i.e., almost identical with $L/L_{0}$ very close to 1). In addition, due to the guidance provided by the flat substrate (the lamellar phase forms perpendicular to the substrate located at one side of the box), all the predictions generated from ML models have very similar lamellar spacing despite the weights of $L_{\Bar{\varphi}}$ loss being used. Thus, taking into account the model performance for volume fraction and flux, for this system, the weight of $1.75$ is adopted throughout the remainder of our analysis.


\subsubsection{Spherical-wall substrate system}
The \textit{average} volume fraction $\Bar{\varphi}_{avg}$ as a function of $L_{\Bar{\varphi}}$ weight is shown by the black curve in Figure~\ref{fig:evaluation}a. Here, the volume fraction is calculated for grid points with density value larger than $0$ (excluding the substrate). From the results, we can see for weights smaller than or equal to $3.0$, the average volume fractions for the predicted trajectories are all very close to $0.5$, showing that all such models obey the correct volume fraction constraint for the spherical-wall case.

The \textit{average} flux as a function of $L_{\Bar{\varphi}}$ weight is shown by the black curve in Figure~\ref{fig:evaluation}b. Weight values of $1.75, 3.0$ and $4.0$ all result in predictions with negligible flux. Although the model with weight $4.0$ gives the smallest flux, it is not desirable because the volume fraction of the model prediction significantly deviates from $0.5$. In contrast, the weights of $1.75$ and $3.0$ result in both volume fraction close to $0.5$ and small amount of flux. Thus, they are candidates for a good model with correct physics.

From the \autoref{fig:spacing} we can see that, again, for most of the weights considered, the ML models give good predictions of lamellar formation with lamellar spacing deviations from simulations of at most $6\%$. Although the weight values of $1.75$ and $3.0$ again both lead to the closest prediction compared with simulations (the $3.0$ weight is even slightly better), to unify our choice of $L_{\Bar{\varphi}}$ weight across three different systems, we still choose the weight for the spherical wall substrate system to be $1.75$ for the rest of our analysis.






\subsection{3D defect dynamics}
The investigation of defect kinetics is crucial for gaining a deeper understanding of the dynamics associated with block copolymer relaxation processes. Advanced sampling techniques have been employed before to study of defect kinetics, but they present several limitations, including their computationally intensive nature and the need for both the initial and final system configurations.
In this study, we have developed a robust \ac{ML} model that addresses these limitations by enabling the accurate prediction of defect kinetics based on the early stages of morphology formation after the spinodal decomposition alone.

Importantly, our model has the capacity to predict long-time kinetics with reduced computational demands, even in the absence of explicit training for this task within the network architecture. We calculate the local nematic order for a 3D lamellar diblock copolymer system and identify defects based on an order threshold. The approach is centered around the calculation of the local nematic order parameter, $S(\mathbf{r})$, at each point in a 3D lattice that comprises a \texttt{64 x 64 x 64} grid with the corresponding local density values, $\varphi(\mathbf{r})$. To achieve this, we first determine the gradient of the local density field, $\nabla\varphi(\mathbf{r})$, utilizing the Sobel operator. Subsequently, we derive the unit normal vector field, $\mathbf{n}(\mathbf{r})$, by normalizing the gradient field. The local nematic order parameter was computed using the orientation tensor, $\mathbf{Q}(\mathbf{r}) = \mathbf{n}(\mathbf{r}) \otimes \mathbf{n}(\mathbf{r}) - \frac{1}{d}\mathbf{I}$, where $\mathbf{I}$ represents the identity matrix, $d$ is the system's dimensionality and $\otimes$ is an outer product operator. To account for thermal fluctuations, we apply a Gaussian filter to the orientation tensor with a standard deviation of $\sigma = 1 R_e$, yielding a smoothed tensor, $\tilde{Q}_{ij}$. The local nematic order parameter was subsequently extracted from tensor $\tilde{Q}_{ij}$.

\begin{figure}[!ht]
\centering
\includegraphics[width=\multifigurewidth]{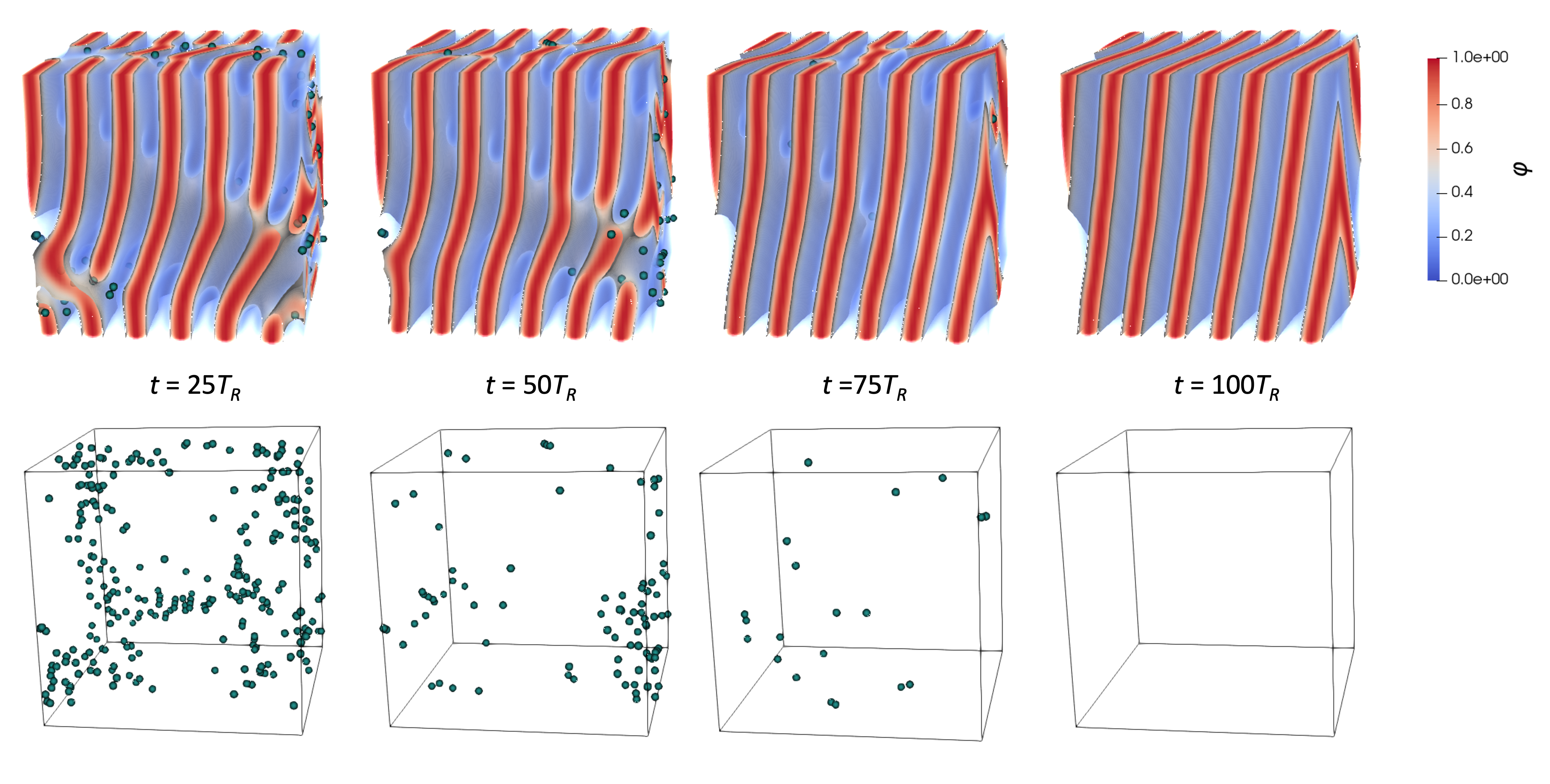}
\caption{
Evolution of predicted lamellar morphology and their defects in a bulk state. The four snapshots correspond to timesteps of 25 $\Tr$, 50 $\Tr$, 75 $\Tr$, and 100 $\Tr$, respectively. Green spheres represent the identified defects. }
\label{fig:defect_bulk_snapshots}
\end{figure}

\begin{figure}[!ht]
\centering
\includegraphics[width=\multifigurewidth]{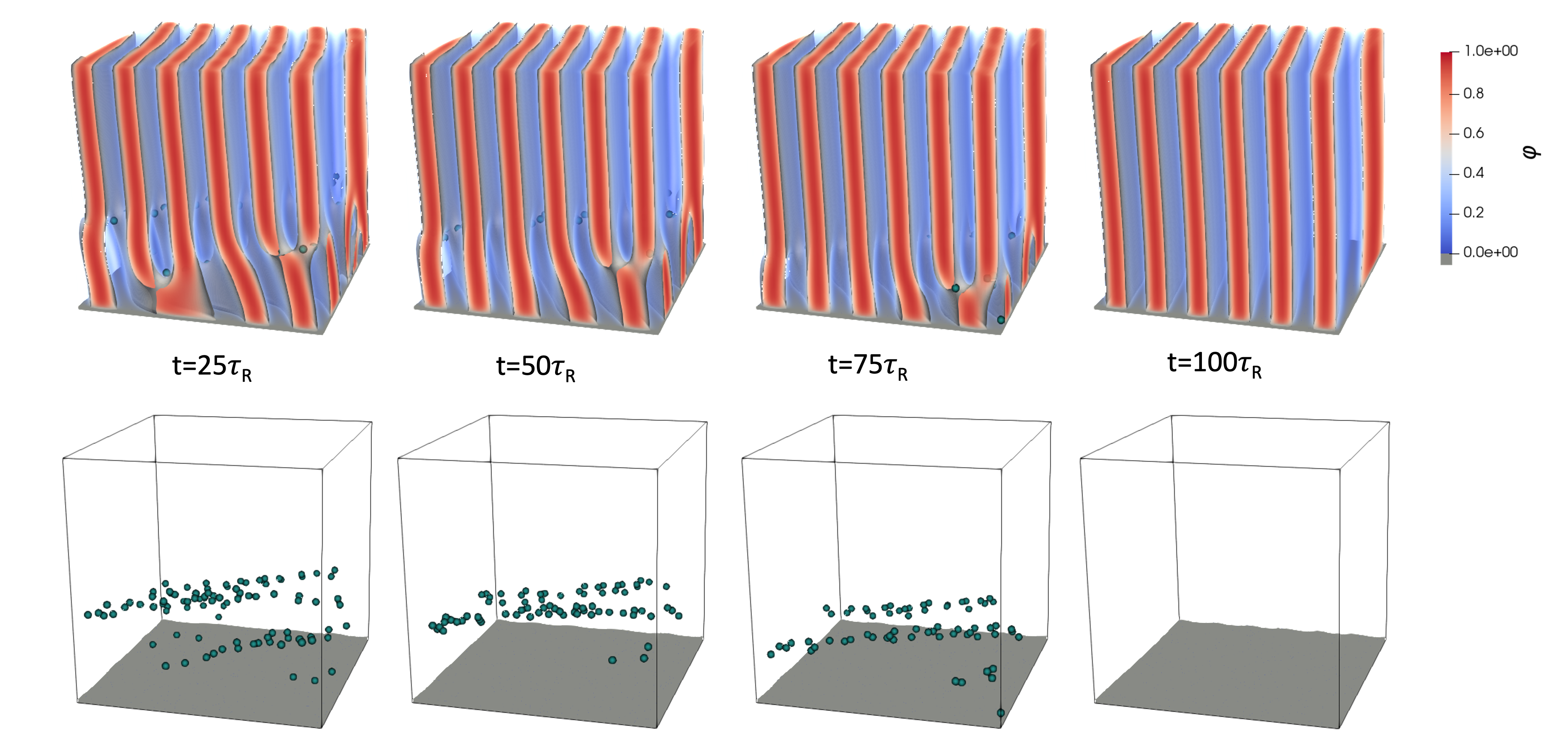}
\caption{
Evolution of predicted lamellar morphology and their defects with flat substrate system. The four snapshots correspond to timesteps of 25 $\Tr$, 50 $\Tr$, 75 $\Tr$, and 100 $\Tr$, respectively. Green spheres represent the identified defects and grey color represents substrate. }
\label{fig:defect_flat_snapshots}
\end{figure}

\begin{figure}[!ht]
\centering
\includegraphics[width=\multifigurewidth]{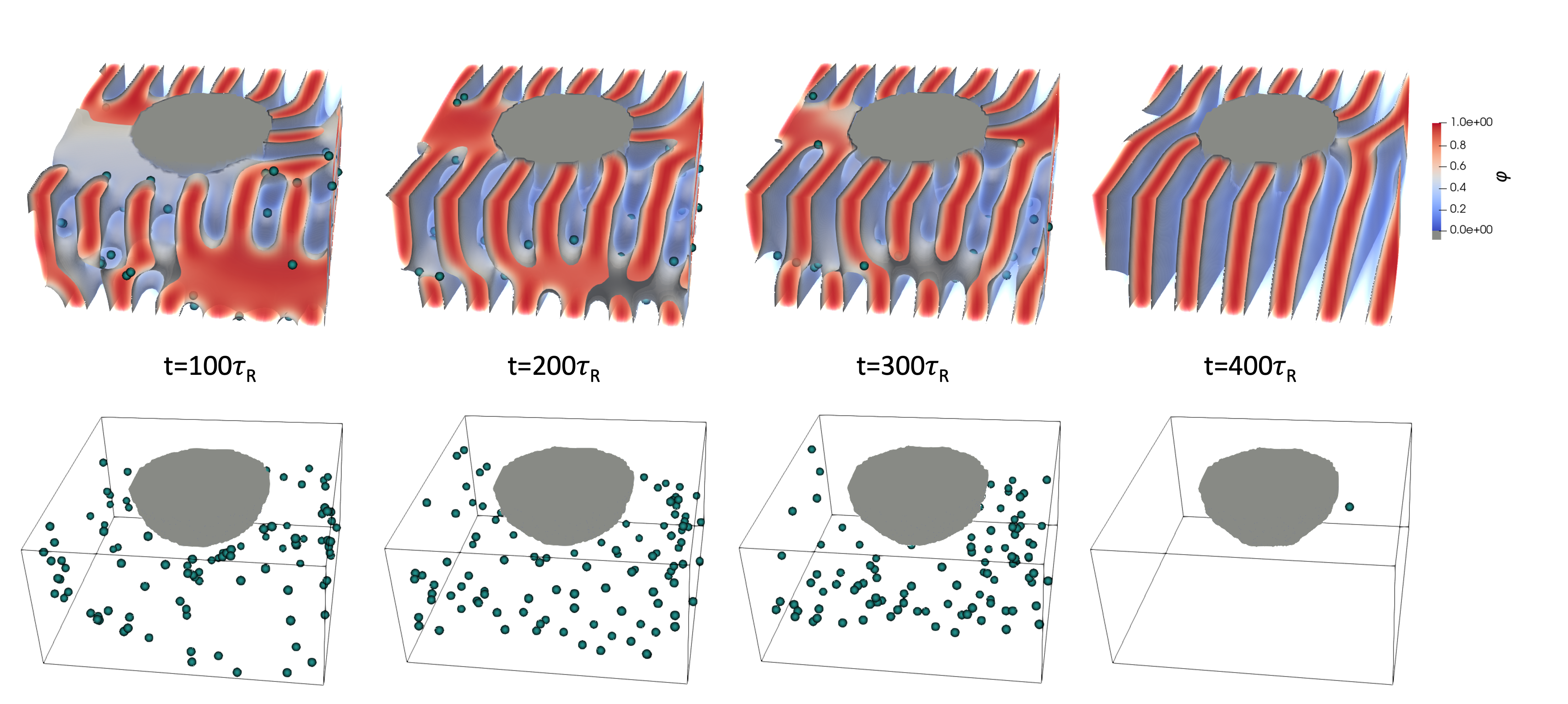}
\caption{Center-cut view snapshots of predicted lamellar morphology and their defects with spherical substrate system. The four snapshots correspond to timesteps of 100 $\Tr$, 200 $\Tr$, 300 $\Tr$, and 400 $\Tr$, respectively. Green spheres represent the identified defects and grey color represents substrate. }
\label{fig:defect_sphere_snapshots}
\end{figure}

\begin{figure}[!ht]
\centering
\includegraphics[width=\multifigurewidth]{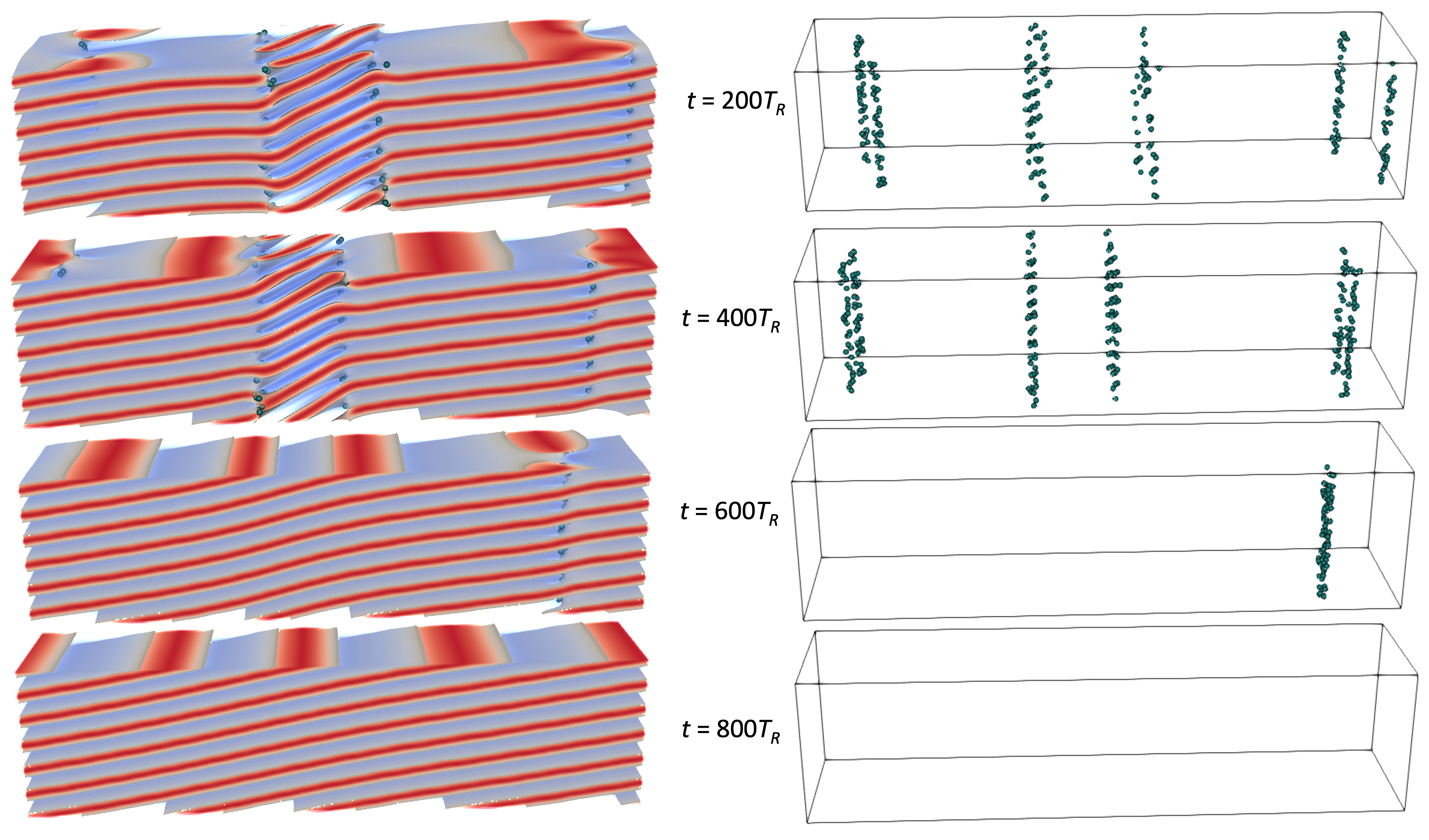}
\caption{Evolution of predicted lamellar morphology and their defects in a bulk state with increasing x-dimensions. The four snapshots correspond to timesteps of 200 $\Tr$, 400 $\Tr$, 600 $\Tr$, and 800 $\Tr$, respectively. Green spheres represent the identified defects.}
\label{fig:defect_bulk_large_snapshots}
\end{figure}
 Three distinct systems were investigated for defect identification: bulk, flat-wall substrate, and spherical wall substrate systems. A sequence of snapshots displaying bulk, flat-wall substrate, spherical wall substrate is depicted in Figure~\ref{fig:defect_bulk_snapshots},Figure~\ref{fig:defect_flat_snapshots}, and Figure~\ref{fig:defect_sphere_snapshots}, respectively. Each figure illustrates the temporal evolution of defects within the systems. Even with varying substrate shapes, our \ac{ML} model accurately identifies defects and annihilates them in a physically plausible manner over extended predicted trajectories. The generated morphologies contain multiple defects, and the \ac{ML} model alters these defects in the same temporal order as the simulation. As demonstrated in Figure\ref{fig:defect_bulk_snapshots}, defects traverse the system and annihilate with opposing defects, resulting in more well-ordered lamellae. Interestingly, Figure~\ref{fig:defect_flat_snapshots} shows that the flat wall substrate system exhibits inherently fewer defects than other systems, which is likely due to the pre-aligned substrate providing a directional influence to facilitate better structuring. On the other hand, Figure~\ref{fig:defect_sphere_snapshots} shows that the defects are slowly decreased due to their geometric curved constraints which have no preferred alignment to one direction. In addition, to better analyze the defect dynamics, we also applied this method on \ac{ML} predictions of a larger bulk system elongated four times along the $z$ axis as shown in Figure~\ref{fig:defect_bulk_large_snapshots}. The \ac{ML} model correctly predicts the evolution of this larger system from random state to the lamellar phase. During this process, we can identify visually many different types of defects and their annealing dynamics such as the elimination of the line defects and the mismatches at the boundaries of two lamellar phases with different orientations.
We note, that the large 3D system exhibits defects, that are pseudo 2D, with a defect propagating perpendicular to the pervading lamellar orientation. This effect can only be studied by this inherent 3D model.
Additionally, we find that these defect are long ranged across the lamellae.
Such defects, have been identified in an earlier study by us in symmetric lamellae of cylindrical confined diblock copolymer melts\cite{schneider2020symmetric}.
In this study, the confinement predominately stabilized these spiraling defects.
Here in a bulk system we observe that these defects predominantly annihilate by collision with other defects.
An intuitive mode of annihilation of defects in 3D outside of grain boundaries, as it allows annihilation of defects independent of their length.
However, we also observe these defects here marking the grain boundary, of the differently oriented grain existing for $t<400\Tr$.
And the grain finally collapses as the defects of opposing grain boundaries collide.
This method of grain annihilation is expected to be slow as the distance traveled of the defects scales with the size of grain.
In this example, we are able to observe it for a small grain spanning less than $10R_{e,0}$ across.

 Overall, over extended time frames, our predictions reveal the complete elimination of defects across all three systems, as evidenced in \autoref{fig:defect_bulk_large_snapshots}. These results serve to confirm that the kinetics of individual defects are accurately ordered and annihilated within our 3D \ac{ML} model.

 We also investigated 2D lamellae with a cylindrical substrate, where we utilized the same methodology as in previous analyses. Defects were identified across a range of predicted morphologies, from $t=0\Tr$ to $t=50000\Tr$, as illustrated in Figure~\ref{fig:2d_defect}. The results provide clear evidence for the accurate detection of defects in the system. As the prediction time progresses, the lamellar morphology exhibits enhanced alignment. This improvement is accompanied by a concurrent evolution and reduction in the number of defects.

\begin{figure}[ht]
\centering
\includegraphics[width=\figurewidth]{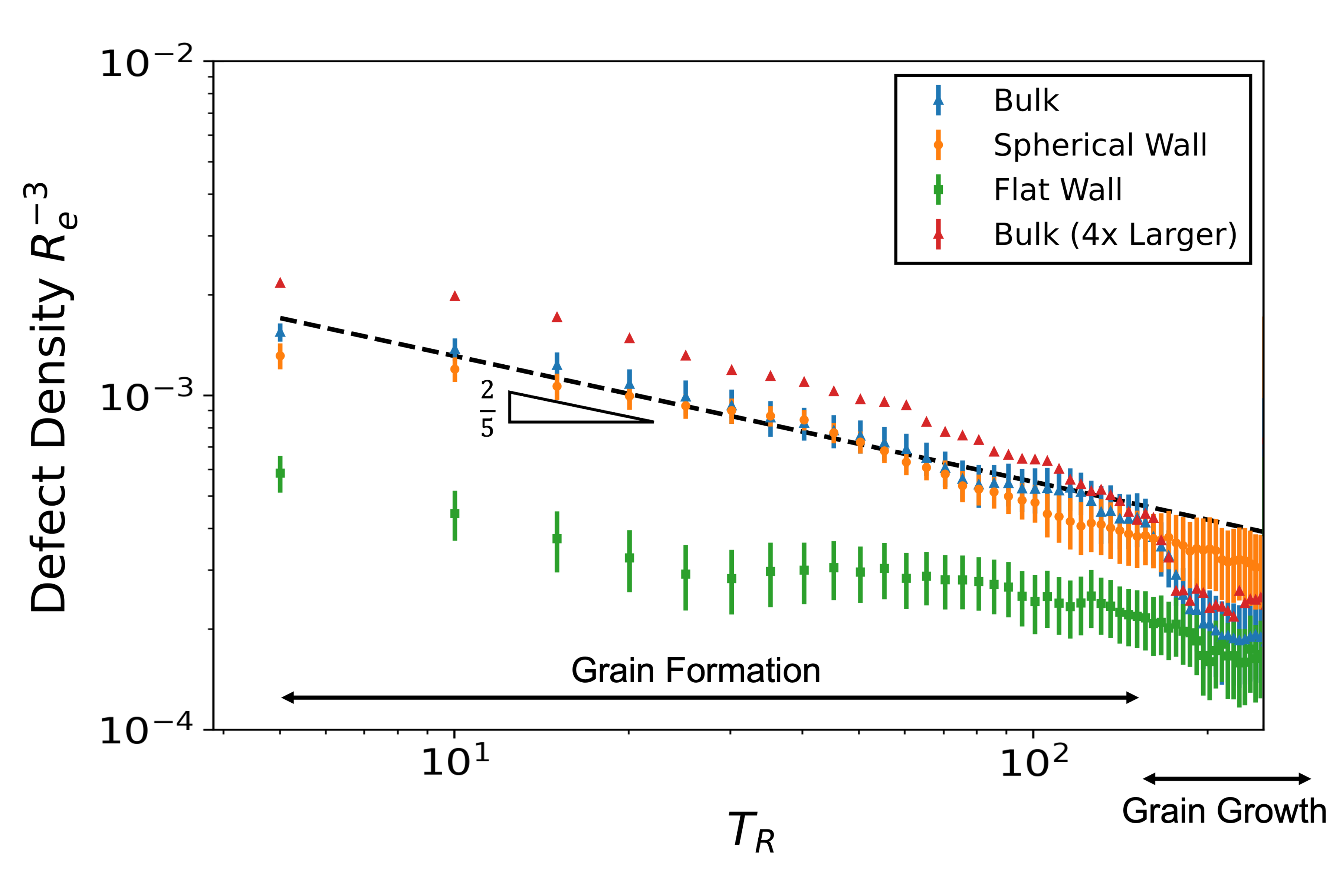}
\caption{ Defect density evolution as a function of time for distinct diblock copolymer systems: bulk, flat-wall substrate, and spherical substrate systems. The large bulk system is also included. The plot exhibits a consistent decrease in defect density over time for all systems, following a power-law behavior during the predicted initial timesteps. As time progresses, the defect density converges, indicating the formation of stable, well-ordered structures in each system. The dashed line represents a power-law fit, with only the fit of bulk system data shown for clarity.}
\label{fig:defect_density}
\end{figure}

Simulation analysis consistently reveals a defect density reduction in all three systems over time, approaching zero as seen in \autoref{fig:defect_density}.
This reduction exhibits a power-law trend, corroborated by experimental findings.
The same trend is evident in the predicted 2D lamellae system, as Figure~\ref{fig:2d_defect_density} depicts.
The consistent slope across decreasing defect densities points to uniform reduction rates.
This is attributed to the relaxation and self-assembly of diblock copolymer chains.
As the systems evolve, the chains reorient to minimize the interfacial area between incompatible blocks, thus reducing defect density.
The initial defect densities in both original and expanded bulk systems (marked by red and blue in Figure~\ref{fig:defect_density}) follow parallel trends, notably decreasing around $150-200 \Tr$ time steps.
This delineates a transition: from a grain formation regime ($5\Tr <t <150-200 \Tr$) characterized by diverse lamellae orientations and interacting defects, to a grain growth regime ($t >150-200 \Tr$) where equilibrium is attained with sparse, weakly-interacting defects.
In the spherical system, the defect density reduction resembles that of bulk systems, suggesting comparable self-assembly dynamics.
This suggests that both defect density and its reduction rate remain consistent irrespective of system size or a spherical substrate's presence.
The latter's effect is likely neutralized by the sphere's intrinsic geometric symmetry, which doesn't favor any particular alignment direction, akin to the spinodal decomposition regime ($t < 5\Tr$) with prevalent defects.
In contrast, the flat-wall system shows fewer defects, influenced by its substrate's asymmetrical constraints that align the copolymer chains.

The results highlight the interaction among system size, substrate geometry, and self-assembly processes in the evolution of diblock copolymer systems. Our \ac{ML} model permits efficient exploration of this relationship while preserving physical accuracy.

\subsubsection{\ac{ML} insights into defect annihilation}\label{sec:xai}
We employ multiple techniques from explainable AI (XAI)\cite{arrieta2020explainable, park2023end} to better understand our  \ac{ML} model. These techniques include saliency maps\cite{simonyan2013deep}, attention maps\cite{oktay2018attention}, Grad CAM, and guided Grad CAM\cite{selvaraju2017grad}. \ac{XAI} is a powerful tool that allows us to explain the output of a blackbox model, without having to understand its underlying functions and architecture. For example, if our image classifier neural network predicts that a picture contains a dog, we can use \ac{XAI} techniques to highlight the parts of the image that contributed to this prediction. Also, an example of  \ac{XAI} application to small molecule applications can be found in Ref.~\cite{park2023end}.

Although \ac{XAI} can be used with any type of model, our \ac{ML} model requires \ac{XAI} techniques that are specifically designed for gradient-based methods or extraction of intermediate tensor information (i.e., \ac{ANN}). One such technique is the saliency map\cite{simonyan2013deep}, which is a gradient of the input with respect to the output. In our work, the output is the $\varphi(r)$ at the next time step, which is a high-dimensional tensor of shape $\texttt{batch x 2 x 64 x 64 x 64}$. This makes it difficult to use the saliency map directly. To address this challenge, we calculate the input gradient with respect to the average volume fraction ($\Bar{\varphi}$) of the predicted output, which is a proxy for the predicted output and has shape $\texttt{batch x 1}$. In other words, we aim to explain which parts of the two previous time step inputs contributed to the predicted output's volume fraction value, resulting in an input gradient of shape $\texttt{batch x 2 x 64 x 64 x 64}$. To visualize this information, we average out the channel size of the saliency map, assuming the difference between two consecutive frame is small.

Attention map\cite{oktay2018attention} is a technique that does not rely on gradients but is a byproduct of our \texttt{\ac{UNet}} during forward propagation.
Each voxel in our dataset has an importance value, which is computed as the output of a sigmoid activation function of a logit obtained by adding the encoder (i.e., key) and decoder (i.e., query or gating) values, as explained in section \nameref{sec:ml}. We collect attention maps generated from different layers of the \texttt{\ac{UNet}} and interpolate them to match the input shape. We then average the maps across the layers to obtain a map of shape $\texttt{batch x 1 x 64 x 64 x 64}$.

Grad CAM \cite{selvaraju2017grad} is a gradient-based method that extracts the gradient of the output from a (decoder) module (e.g. convolution, fully-connected, etc.) and computes the spatial average value per channel (i.e., $\texttt{batch x channels x 1 x 1 x 1}$). This value is then multiplied with the output from the same module (i.e., $\texttt{batch x channels x dim1 x dim2 x dim3}$) and averaged across the channel dimension, resulting in a tensor of shape $\texttt{batch x 1 x dim1 x dim2 x dim3}$. We then interpolate the spatial dimensions and collect the interpolated output from all decoder layers to average it out across layers, resulting in a Grad CAM output of shape $\texttt{batch x 1 x 64 x 64 x 64}$, where $\texttt{channels}$ refers to the hidden channel and $\texttt{dim1, dim2, dim3}$ refer to the convolved spatial dimensions. The \texttt{dim1}, \texttt{dim2} and \texttt{dim3} are less than \texttt{64}.

We employ various \ac{XAI} techniques to explain our \ac{ML} model's predictions and understand its decision-making process. Specifically, we use saliency maps and attention maps\cite{arrieta2020explainable, park2023end, simonyan2013deep, oktay2018attention, selvaraju2017grad, springenberg2014striving}, since our internal testing showed that Grad CAM does not capture patterns as well as saliency maps or attention maps. Nevertheless, these techniques allow us to gain insights into why and how the model makes certain predictions, even though we do not know the exact functions and architecture of the model.

The different \ac{XAI} methods offer different insights into the underlying data of morphology evolution. The attention map is particularly useful in \textit{capturing} defects, while the saliency map highlights the regions of the system that our \ac{ML} model will \textit{eliminate} in future timesteps. To demonstrate the effectiveness of these methods, we apply our \ac{ML} model trained on simple 3D bulk system data to predict the evolution process of a special case where two lamellar phases with different orientations coexist in the simulation. One lamellar phase is aligned along the $xy$ plane, while the other is aligned along the $yz$ plane.
According to particle-based simulations, the lamellar phase aligned along the $yz$ plane will propagate through the system, while the $xy$ plane-oriented lamellar phase will disappear. The orientation of the morphologies in space is arbitrary and only exhibited by this specific sample.
If our assumption is correct, the \ac{ML} model should be able to predict this evolution process. In this case, the attention map will highlight the defect locations shown in the snapshot of the system, while the saliency map will highlight the region that will disappear, such as the horizontal lamellar phase.

\begin{figure}[ht]
\centering
\includegraphics[width=\multifigurewidth]{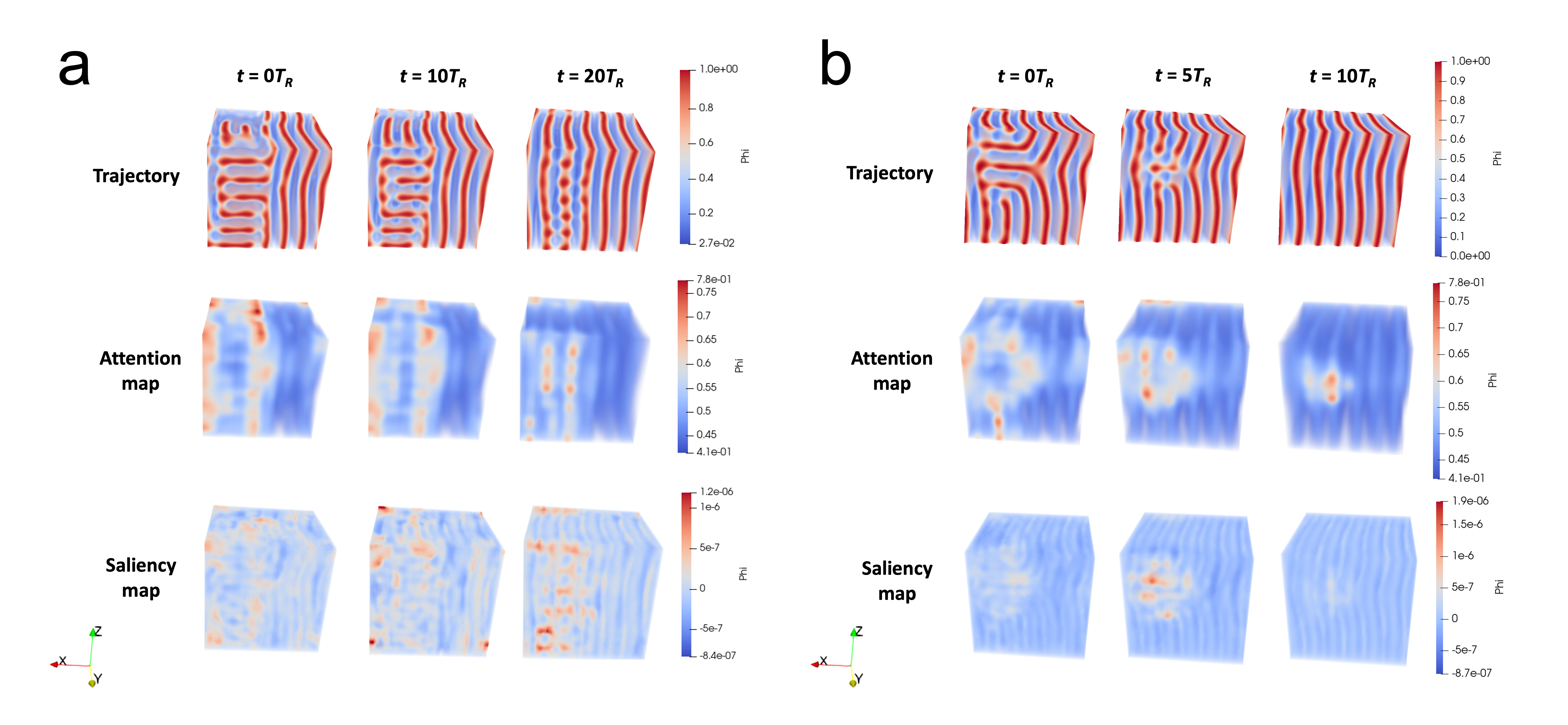}
\caption{Trajectory, attention map, and saliency map generated by ML model using simulation snapshots of different timesteps: a) early stage; b) latter stage. The \ac{ML} model is trained with simple 3D bulk system data and during the inference, predictions are made for snapshots chosen from a special simulation system where two lamellar phases with different orientation coexist.}
\label{fig:xai}
\end{figure}

The ML-generated trajectory in Figure~\ref{fig:xai}a shows that although the model was not trained with data of this special case, it is able to be generalized to this system and successfully predicts the formation of the lamellar phase along the $yz$ plane as verified by the actual simulation. The attention map highlights the boundaries of these two mismatched lamellar phases as shown by the red spots in the snapshots of timestep $0$ and $2$. In timestep $4$, the lamellar phase parallel to the $yz$ plane already dominates the system while some dislocations still exist. The attention map at timestep $4$ highlights those defects with red spots. While the attention map mainly shows large values at certain spots correlated to defects, the saliency map tends to be more smeared and highlights regions to be eliminated. For example, in timestep $4$ of Figure~\ref{fig:xai}a, the saliency map highlights the left half of the box which is transitioning from $xy$ plane-oriented to $yz$ plane-oriented.

Similar observations are also made for Figure~\ref{fig:xai}b where the predictions are generated from later stage simulation snapshots. In this case, part of the lamellar phase at the left half of the box has already aligned to the $yz$ plane. The model again correctly predicts the elimination of line defects and the formation of lamellar phase parallel to the $yz$ plane. During this process, we can also see that the highlights in the attention map correlate well with defects such as dislocations and line ends. Similar to the previous case, while the attention map has strong point-like signals at the defective locations and boundaries, the saliency map tends to highlight the entire region of interest. For example, in time step $0$, the saliency map has larger values at the upper left corner of the box where most of the defects exist and are about to be eliminated. In the meantime, the attention map shows red spots at the boundary where the defects are located. From this preliminary effort, we demonstrate that \ac{XAI} techniques can be a valuable tool to decipher the convolution-based \ac{ML} model and shed light on the understanding of how the \ac{ML} model decides to evolve the system.

\subsection{Diffusion analysis}
Diblock copolymers hold significant potential for application as electrolytes in solid-state batteries, which are intrinsically safer and represent the next generation of high-performance energy storage solutions. Diblock copolymer materials possess the unique ability to concurrently optimize properties that are typically mutually exclusive, such as mechanical stability and ionic conductivity\cite{morris2017harnessing}. Nevertheless, the comparatively low conductivity of diblock copolymer materials presents a challenge for the continued advancement of solid-state batteries. Given that ionic conductivity arises from charge transport processes, it is imperative to investigate the diffusion processes in block copolymer electrolytes, which are fundamentally reliant on the \ac{BCP} nanostructure\cite{shen2018diffusion,schneider2019engineering}. In previous work we have investigated diffusion during the formation of microphase separation\cite{schneider2019engineering}. But the time and length scales involved in this process have  prevented in-depth analysis for different situations.

By implementing our \ac{ML} model this computational limitation can be overcome. Now, we are able to predict the kinetics of \ac{BCP} morphology evolution, particularly over extended time scales, allowing for an exploration of the impact of spatial orientation and morphology evolution on the diffusion process.

\begin{figure}[!ht]
\centering
\includegraphics[width=\multifigurewidth]{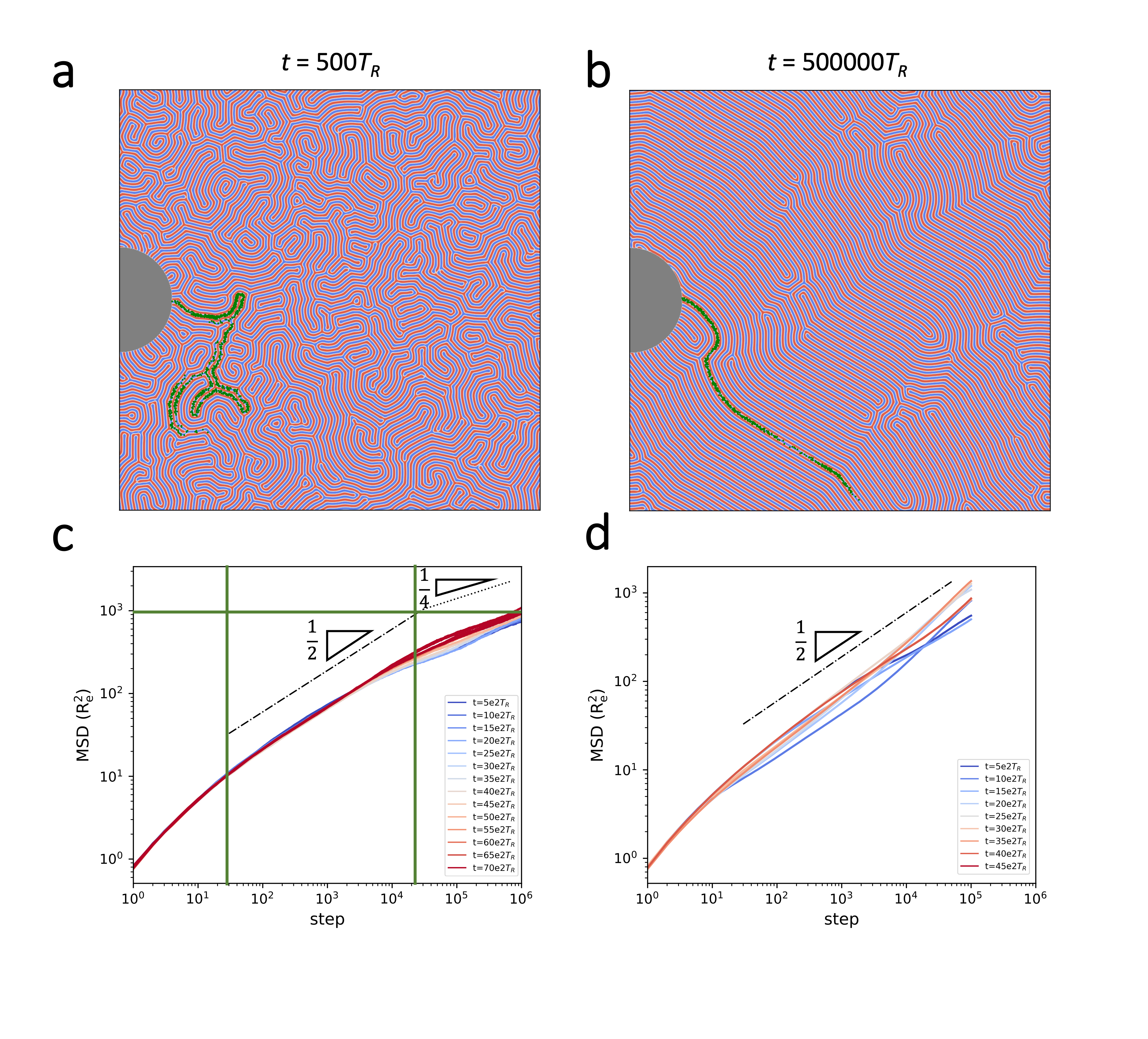}
\caption{Diffusion dynamics in two-dimensional systems with different morphologies are:
(a) Diffusion in short ML-predicted morphology at $ t = 5e2 \Tr $.
(b) Diffusion in elongated ML-predicted morphology at $ t = 5e5 \Tr $.
(c) MSD trajectories for short ML-predicted morphology highlight an initial subdiffusive phase with a $ \frac{1}{2} $ slope, transitioning to a fractal diffusion phase with a $ \frac{1}{4} $ slope.
(d) MSD trajectories in the elongated ML-predicted morphology consistently exhibit subdiffusive motion with a $ \frac{1}{2} $ slope, attributable to reduced defects.
The x-axis denotes random walk \ac{MC} steps, the y-axis represents \ac{MSD}, and color differentiation corresponds to morphology evolution time.}
\label{fig:diffusion}
\end{figure}

As an initial step, prior to addressing more intricate structures, we consider the diffusion process in a two-dimensional system employing a spherical substrate. Our analysis involves the examination of a simple random walk model, wherein a single particle is limited to discrete movements of one $R_e$ distance at a time, and is restricted to diffusing within one domain, with diffusion through the alternate domain prohibited.

We applied the \ac{ML}-predicted morphology across a temporal span from 500 ($\Tr$) to 7000 ($\Tr$).
A specific snapshot showing the morphology we diffuse in at $ t=500 \Tr $ is depicted in Figure~\ref{fig:diffusion}a.
Upon concluding 100,000 steps, ($\mathbf{s}$), we contrast the \ac{MSD} across different \ac{ML}-predicted morphologies, showcased in Figure~\ref{fig:diffusion}c.
The log-log \ac{MSD} plot delineates two regimes indicative of unique diffusion behaviors.
Initially, for $\mathbf{s}=25000$ steps, the congruent \ac{MSD} curves have a 1/2 slope, signifying a subdiffusion phase.
This is attributed to diffusion hindrance by neighboring grains forming. Beyond these initial steps, the \ac{MSD} curves diverge.
The \ac{MSD} for the longer-duration predicted morphology (t=7000 $\Tr$) surpasses its counterpart at $ t=500 \Tr $, consistent with the notion that a more structured, extended-duration morphology augments diffusion.
In the subsequent phase, from $\mathbf{s}=25000$ to $\mathbf{s}=1000000$, the \ac{MSD} curves exhibit a 1/4 slope on a log-log scale.
This attenuated diffusion is characteristic of constrained systems or heterogeneous settings, like fractals or disordered mediums, where diffusion dynamics are sculpted by the medium's architecture. This "fractal diffusion" paradigm~\cite{gmachowski2015fractal} elucidates particle motion within intricate fractal constructs, where the medium's complexity confines diffusion, leading to the noted 1/4 exponent in the \ac{MSD} scaling.

Particle diffusions were analyzed for the longer ML-predicted morphologies over a time span from $ 5\cdot10^4 \Tr $ to $ 5\cdot10^5 \Tr $.
A specific snapshot of the long \ac{ML}-predicted morphology at $ t=5\cdot10^5 \Tr $ is illustrated in \autoref{fig:diffusion}b.
Concurrently, the \ac{MSD} for different \ac{ML}-predicted morphologies is detailed in \autoref{fig:diffusion}d.
Notably, the extended prediction time facilitates a highly ordered lamellar structure, reducing defects and the presence of fractal configurations.
Over $ 10^6 $ diffusion steps, the system consistently exhibits a subdiffusive behavior due to obstruction from defects in the morphology, evidenced by a 1/2 slope in Figure~\ref{fig:diffusion}d. The lack of a 1/4 slope confirms the absence of fractal diffusion within this range, with the interpretation that once grains are formed, the diffusion happens within a grain, which has a well-ordered morphology, not a fractal with defects anymore.

For systems during the grain forming phase at $\mathbf{s}=500\Tr$, diffusion manifests in two distinct regions. Initially, we observe a subdiffusive behavior with a $1/2$ exponent as particles move within individual lamellae. However, as diffusion continues, the numerous defects between the lamellae further impede particle movement, leading to diffusion patterns reminiscent of fractals, as depicted in \autoref{fig:diffusion} a\&c.

In contrast, the diffusion behavior in a well-developed lamellar structure at $\mathbf{s}=500\,000\Tr$ shows a different pattern. In the absence of defects, the diffusion strictly follows a $\propto t^{1/2}$ trend, constrained only by the predominantly linear lamellae, even over extended durations, as depicted in \autoref{fig:diffusion} b\&d.

\section{Conclusions}

We present an \ac{ML} approach that accelerates the computation of morphology evolution in large time scales, utilizing separated time scales between the particle evolution and the slow morphology evolution.
This approach overcomes the limitations of empirical continuum models by learning stochastically driven defect annihilation directly from particle-based simulations.
Building on previous work\cite{schneider2021combining}, a new focus on full three-dimensional systems with complex boundary conditions and confinements required a new \ac{ML} model based on the \texttt{\ac{UNet}} architecture. The validity of the model is ensured through the introduction of physical concepts via the loss function and symmetries via data augmentation.
The model is shown to achieve fundamental physical correctness in mass, momentum, and micro-structure conservation, and is validated for three separate use cases.
The ability to generate large system sizes and long trajectories allows for systematic investigation of defect densities under different confinements, enabling the study of morphology evolution and resulting final structure as a function of confining geometry and system size.
The \ac{ML} model also facilitates the use of \ac{XAI} methods to understand and visualize how the \ac{ML} methods evolve the morphology in time, allowing for validation with physical intuition.
By applying this capability to the study of diffusion of particles inside a single block in late-stage micro-morphologies, it is observed that the diffusion characteristics fundamentally change at late times, highlighting the importance of access to late-stage morphologies for the understanding of this process.
Overall, the proposed \ac{ML} approach represents a significant advancement in our ability to study morphology evolution in complex systems, providing a powerful tool for future research in this field.

\section*{Acknowledgments}
This work is supported by NIST, through the Center for Hierarchical Materials Design (CHiMaD).
This research used resources of the Argonne Leadership Computing Facility, which is a DOE Office of Science User Facility supported under Contract DE-AC02-06CH11357.
We acknowledge the University of Chicago Research Computing Center for support of this work.
We thank Eliu Huerta, Tom Gibbs, Pablo Zubieta, L.S., Joshua Mysona and J.P. for organizing an Argonne National Lab and Nvidia sponsored GPU-Hackathon, where this team was formed and the project idea formed.
We thank Marcus M\"uller for fruitful discussions.
L. Schneider gratefully acknowledges the support of the Eric and Wendy Schmidt AI in Science Postdoctoral Fellowship, a Schmidt Futures program.


\section*{Data and code availability}
The \ac{ML} models presented in this article
are open source in GitHub \url{https://github.com/gsun33/3D_ML/tree/main/Files_from_AIHacks/TEAM_HyunBoyuan/AIHack2022-main}. Following best practices, the model checkpoints and datasets used in our work are published in Zenodo \url{https://zenodo.org/record/7996198}.

\break

\bibliography{main}


\break

\section{Supplemental Materials}

\setcounter{figure}{0}
\renewcommand{\thefigure}{S\arabic{figure}}

\begin{figure}[ht]
\centering
\includegraphics[width=\linewidth]{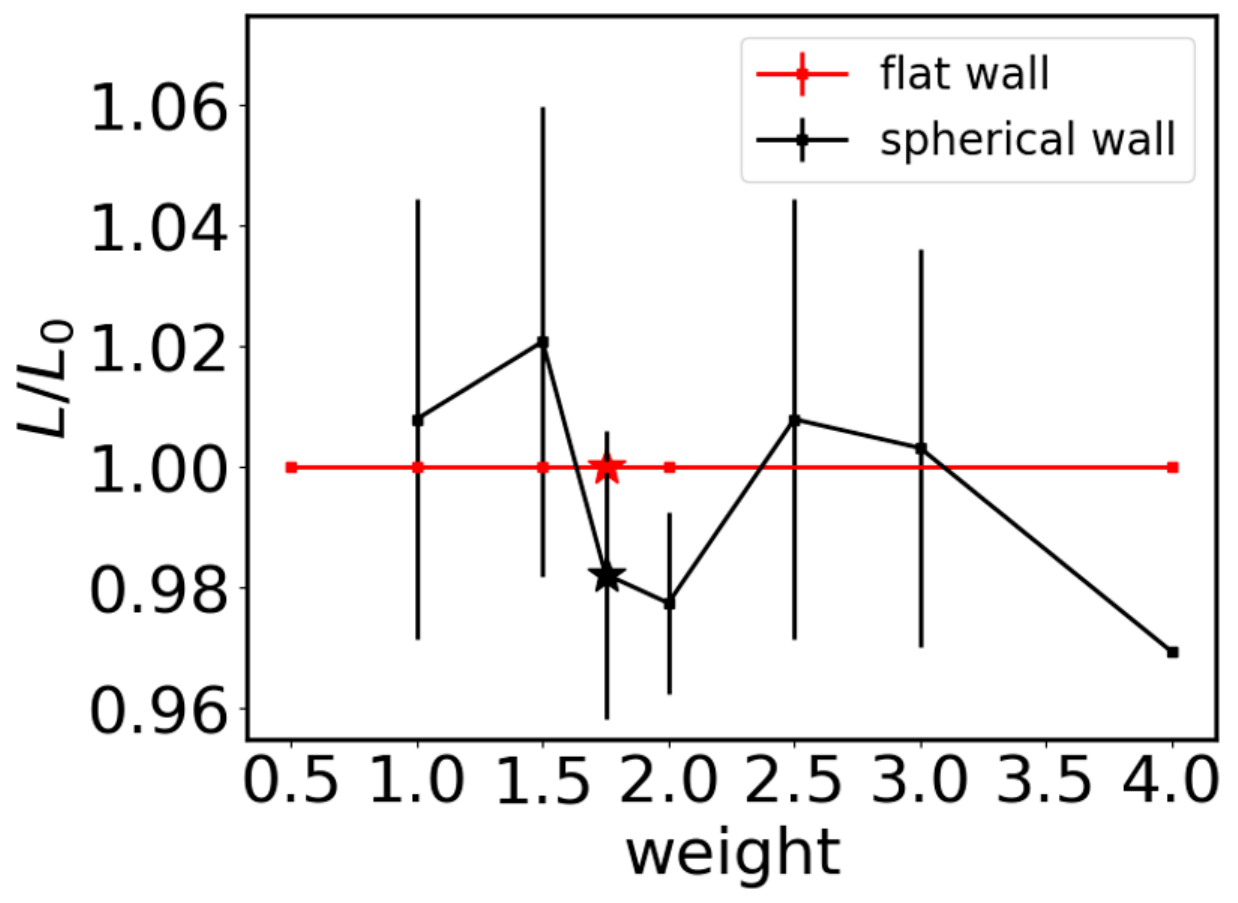}
\caption{Lamellar spacing $L$ scaled by the natural lamellar spacing $L_{0}$ determined by simulation as a function of $L_{\texttt{volume fraction}}$ weight for flat wall substrate and spherical wall substrate systems. The star symbols corresponds to the weight of $L_{\texttt{volume fraction}}$ being chosen for further analysis.}
\label{fig:spacing}
\end{figure}

\begin{figure}[ht]
\centering
\includegraphics[width=0.85\linewidth]{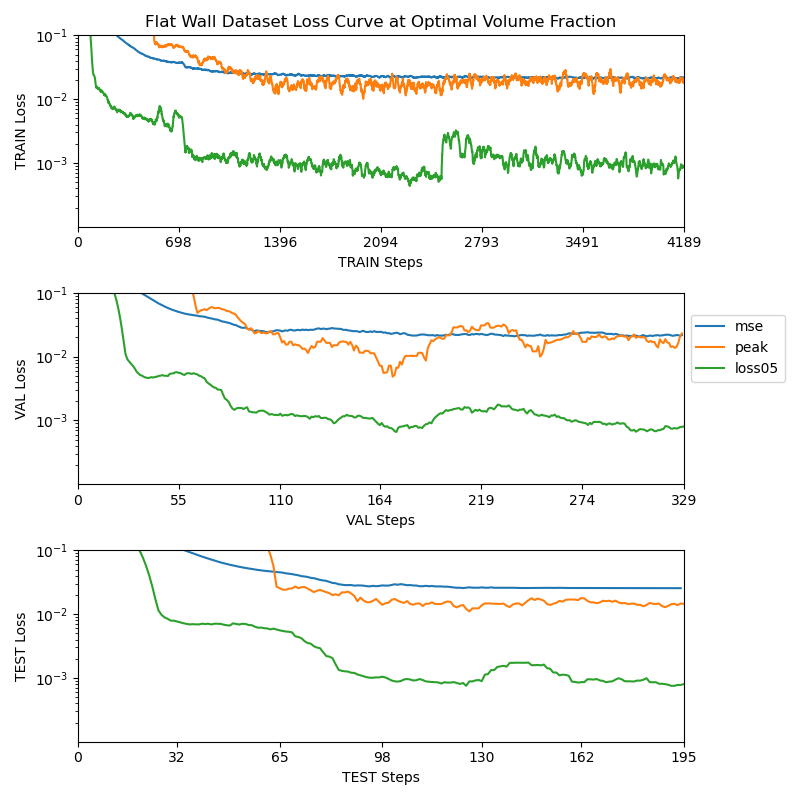}
\caption{The loss values over training/validation/test steps for fall wall dataset (labeled original dataset). The top, middle and bottom panel is train, validation and test loss curve, respectively. There are three losses used: \textit{mse} (blue), \textit{peak} (orange) and \textit{volume fraction} (green) losses. Training steps are batch iteration steps during training to optimize our \texttt{UNet} model. The validation steps are batch iteration steps during validation to save better performing (i.e. validation loss drop) models. Lastly, the testing steps are batch iteration steps tested after all the training and saving are over. The result is logged at every 5th step of iteration. }
\label{fig:flat_error}
\end{figure}

\begin{figure}[ht]
\centering
\includegraphics[width=0.85\linewidth]{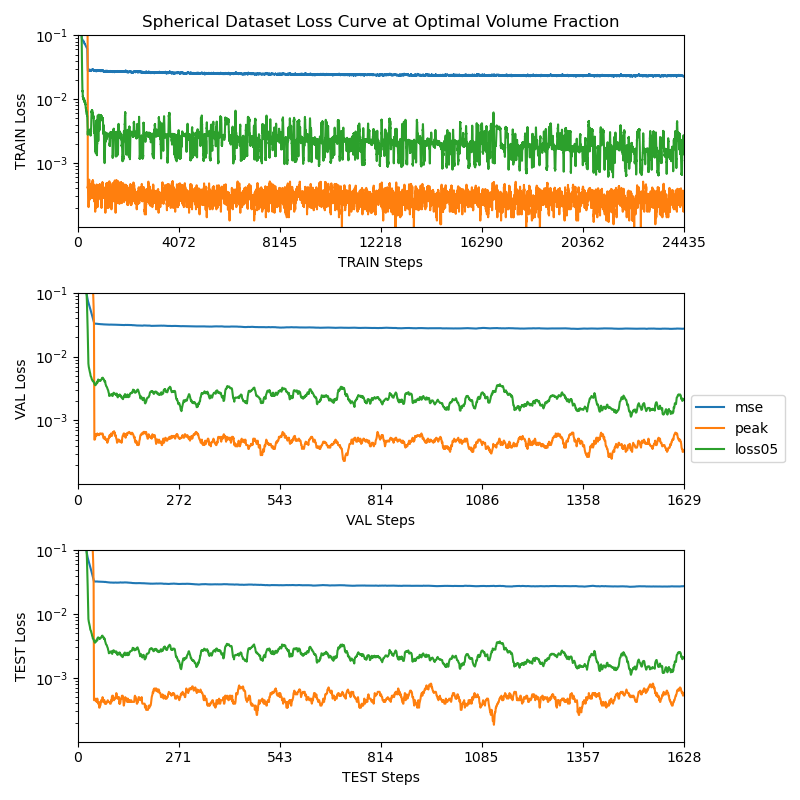}
\caption{The loss values over training/validation/test steps for spherical wall dataset (labeled original dataset). The top, middle and bottom panel is train, validation and test loss curve, respectively. There are three losses used: \textit{mse} (blue), \textit{peak} (orange) and \textit{volume fraction} (green) losses. Training steps are batch iteration steps during training to optimize our \texttt{UNet} model. The validation steps are batch iteration steps during validation to save better performing (i.e. validation loss drop) models. Lastly, the testing steps are batch iteration steps tested after all the training and saving are over. The result is logged at every step of iteration. }
\label{fig:sph_error}
\end{figure}

\begin{figure}[ht]
\centering
\includegraphics[width=0.85\linewidth]{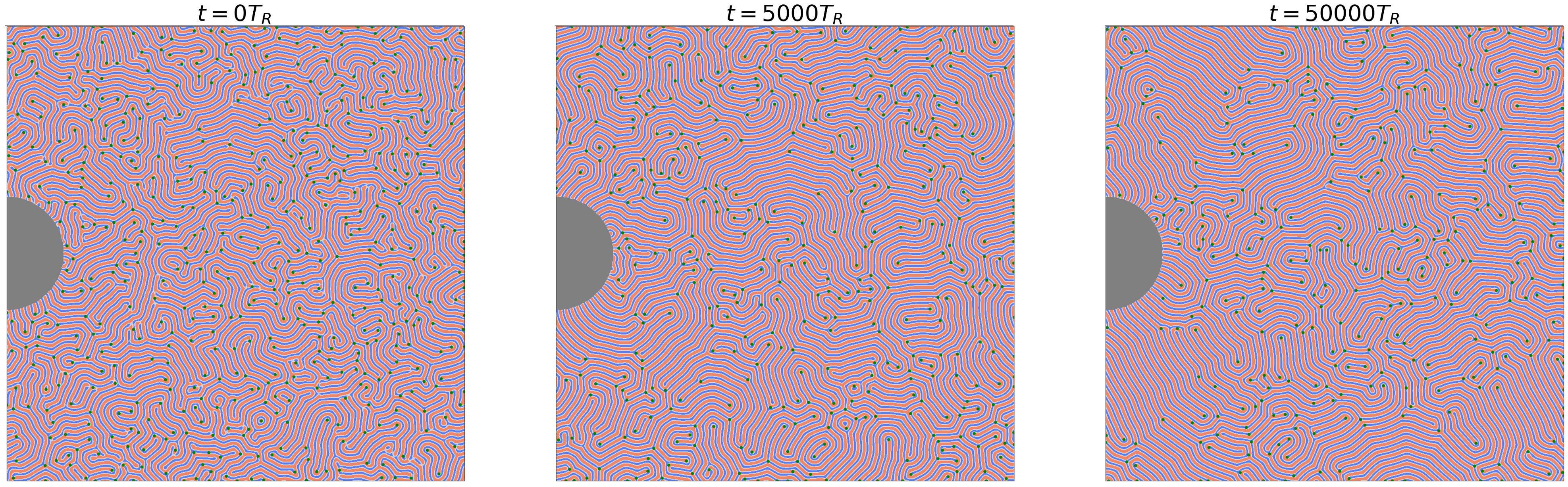}
\caption{Evolution of defects in 2D lamellae with a cylindrical substrate. The evolution of defects is over a range of predicted morphologies, from $t=0T_R$ to $t=50000T_R$, demonstrating the accurate identification of defects and their subsequent reduction as the lamellar morphology becomes more aligned over time.}
\label{fig:2d_defect}
\end{figure}

\begin{figure}[ht]
\centering
\includegraphics[width=0.85\linewidth]{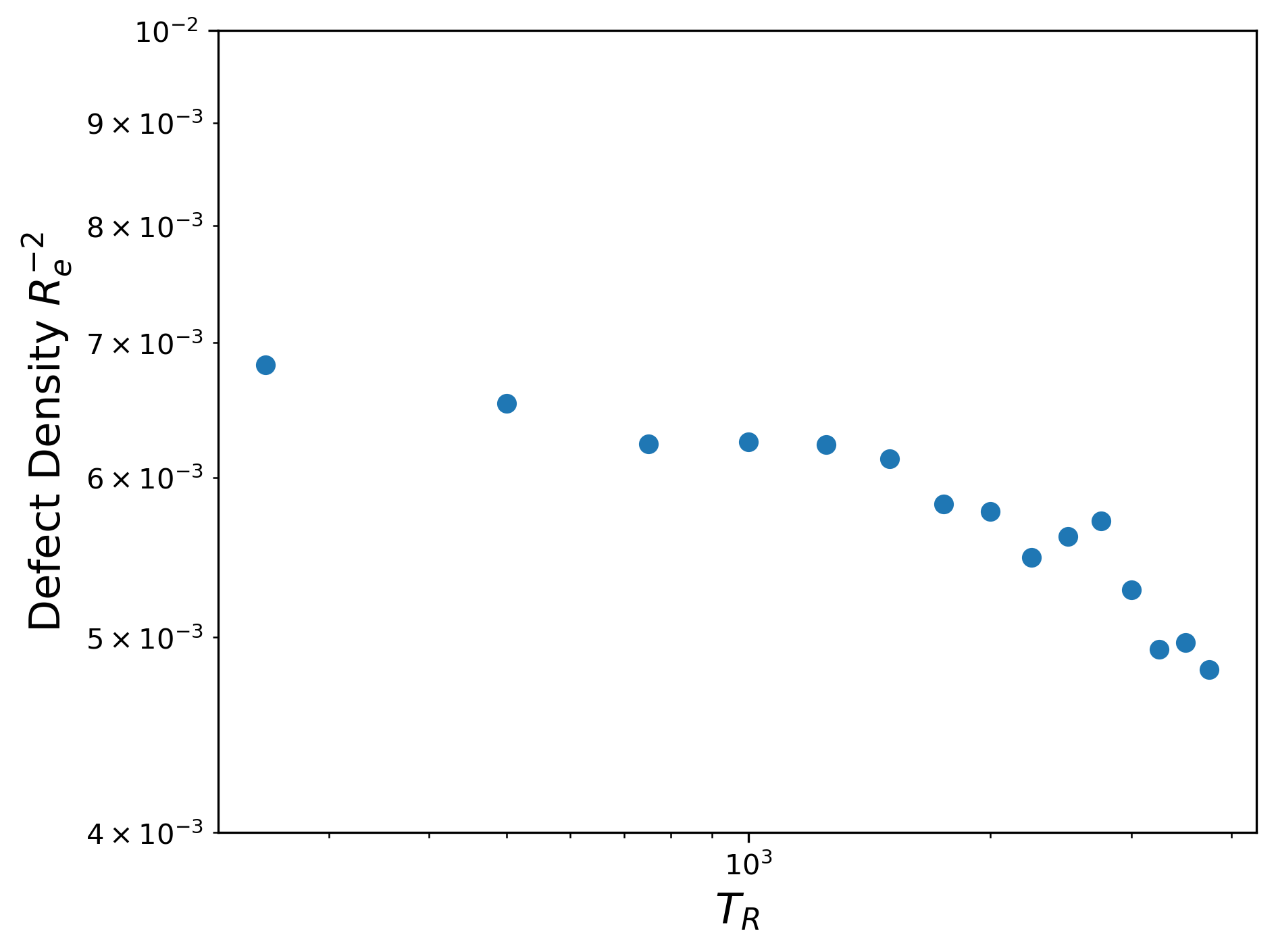}
\caption{Defect density in predicted 2D lamellae with a cylindrical substrate. It follows a power-law behavior as in 3D cases.}
\label{fig:2d_defect_density}
\end{figure}






\end{document}